\documentclass[12pt]{article}
\usepackage{graphicx}  
\usepackage{dcolumn}   
\usepackage{bm}        
\usepackage{amsmath}
\usepackage{amssymb}   
\usepackage{xcolor}
\usepackage{subcaption}
\usepackage[title]{appendix}

\usepackage{hyperref}
\begin{document}

\baselineskip=24pt

\begin{titlepage}

\centerline{\bf Modeling Temperature Profiles in the Pedestal of NSTX with Reduced Models}

\bigskip

\centerline{P.-Y. Li$^{1}$, D. R. Hatch$^{1,2}$, L. A. Leppin$^{1,3}$, J. Schmidt$^{1}$, J. F. Parisi$^4$, M. Lampert$^{4}$,}
\centerline{ M. Kotschenreuther$^2$, and S. M. Mahajan$^{1,2}$}

\medskip

 \centerline{\it $^1$Institute for Fusion Studies, The University of Texas at Austin, Austin, TX, USA.}
 \centerline{\it $^2$ExoFusion, Austin, TX, USA.}
 \centerline{\it $^3$Oden Institute for Computational Engineering and Sciences, Austin, TX, USA.}
 \centerline{\it $^4$Princeton Plasma Physics Laboratory, Princeton, NJ, USA}

\vspace{0.25cm}

\begin{abstract}
\baselineskip=24pt
This paper describes new modeling capabilities for predicting H-mode pedestal profiles in spherical tokamaks. Temperature profiles for NSTX discharges 132543 and 132588 are modeled by coupling the \textsc{astra} transport solver with neoclassical transport and gyrokinetic-based reduced models for electron temperature gradient (ETG) and kinetic ballooning mode (KBM) instabilities. A quasi-linear surrogate model for ion-scale transport is developed using linear \textsc{gene} simulations, requiring only a single free parameter calibrated to one discharge. Time-evolving the temperatures with fixed density yields good agreement with experiments for both discharges. Systematic analysis of the transport mechanisms reveals that neoclassical transport is huge across the entire pedestal region for the ion channel. ETG turbulence is large in the plasma edge and low density gradient region, contributing substantially to the electron channel. However, KBM/MHD-like modes also drive significant transport in both the ion and electron thermal channels, making them essential for accurate pedestal modeling. Further refinements, including explicit $E \times B$ shear suppression and scaled ETG transport, produce quantitative but not qualitative improvements. This work lays the foundation for predictive modeling of future devices.
\end{abstract}
\end{titlepage}

\section{Introduction}

Accurately predicting pedestal temperature profiles is essential for understanding and forecasting the performance of magnetically confined fusion plasmas. In this work, we demonstrate a new modeling capability for predicting pedestal temperature profiles by integrating gyrokinetic-based reduced models for pedestal transport into a time-dependent transport framework, and apply this approach to two representative NSTX discharges. Similar predictive capabilities are currently being validated for DIII-D and broader standard aspect ratio scenarios.

The transport in the H-mode pedestal results from a complex interaction of multiple physical mechanisms that operate across a wide range of spatial and temporal scales~\cite{Doyle2007,Howard2016}. Predicting pedestal profiles in spherical tokamaks has proven challenging. While some recent approaches have attempted to predict the pedestal using only kinetic ballooning mode (KBM) constraints or adaptations of the EPED model~\cite{Parisi2024}, a fully established and broadly validated capability does not yet exist. While macroscopic MHD instabilities, such as peeling-ballooning modes, often define the ultimate global limits of the pedestal, turbulence driven by the electron temperature gradient~\cite{Horton1999,Jenko2000,Dannert2005,Jenko2005} (ETG) instability is frequently identified as a major contributor to the electron thermal channel. It is now increasingly recognized that ETG turbulence can play an important role in regulating the electron temperature pedestal well before macroscopic MHD limits are reached~\cite{Hatch2017,Kotschenreuther2019,Li2020,Groebner2022,Parisi2023}. Similar to standard aspect ratio pedestals, transport from microtearing modes (MTM) is also thought to contribute in the pedestal~\cite{Guttenfelder2012}. Kinetic ballooning modes (KBM) are also predicted to contribute in all transport channels and constrain the pedestal pressure~\cite{Diallo2013, Kaye2013}.    

It is essential to highlight that comprehensive transport modeling—specifically accounting for the dynamic thermal exchange between species—is required for a full understanding of pedestal transport and dynamics. In this work, linear and nonlinear gyrokinetic simulations are performed using the \textsc{gene} code~\cite{Jenko2000, Gorler2011}. We model $T_e$ and $T_i$ profiles in the NSTX pedestal using the \textsc{astra} integrated modeling suite~\cite{Pereverzev2002, Fable2013, Tardini2026} by fixing the density profiles and evolving temperature profiles in response to various transport mechanisms. Modeling density profiles is challenging due to, among other things, uncertainties in the particle source. Consequently, we focus here on the thermal channels and leave density evolution for future work. We test different transport mechanisms systematically through a step-wise progression in order to understand their impact on the pedestal. We begin by exploring the role of ETG turbulence in isolation using a recently developed reduced ETG transport model~\cite{Hatch2024}. This step serves as a useful thought experiment to assess how far ETG-driven transport alone can account for the observed pedestal structure in NSTX. Perhaps unsurprisingly, additional transport mechanisms are necessary to reproduce experimental pedestal profiles. Motivated by these considerations, we then extend the modeling framework by incorporating gyrokinetic-based reduced models for kinetic ballooning mode (KBM) and MHD-like transport, as both ETG turbulence and ion-scale KBM/MHD-like modes are frequently observed to play critical roles in the pedestal region.

The modeling framework is applied to two representative NSTX discharges with contrasting pedestal characteristics. Discharge 132543 is an ELM-free, lithium-conditioned plasma exhibiting a relatively wide pedestal, while discharge 132588 is a non-lithiated, ELMy discharge with a comparatively narrow pedestal. By applying the reduced transport models within the \textsc{astra} transport solver~\cite{Pereverzev2002}, we dynamically evolve the pedestal temperature profiles and evaluate the relative roles of ETG, neoclassical, and ion-scale MHD-like transport mechanisms. We find that neoclassical transport is huge across the whole pedestal region for the ion transport channel. ETG is large in the plasma edge and low density gradient region, which contributes substantially to the electron channel. Furthermore, KBM/MHD-like modes are also large and contribute to both ion and electron channels. The quasilinear model has only one free parameter, which was tuned so that the simulations reproduce the experimental profiles in one of the discharges.  All other aspects of the model are fully determined by the gyrokinetic-based ETG model, the quasilinear treatment of the ion-scale linear gyrokinetic simulations, and neoclassical transport.  With this single free parameter, the simulations reproduce with good agreement the $T_e$ and $T_i$ profiles from the pedestal top to the separatrix for the two NSTX discharges.  These models provide a new capability for efficiently capturing key pedestal transport phenomena. It lays the foundation for more extensive validation, more comprehensive modeling (including density profile prediction), and ultimately predictive modeling, design, and optimization of future fusion devices.

The paper is organized as follows. Section~\ref{sec:etg_neo} introduces the reduced ETG transport model and compares its modeled heat fluxes with nonlinear gyrokinetic simulations from \textsc{gene} and \textsc{cgyro}~\cite{Candy2016}. Using the \textsc{astra} framework, we then evaluate the predicted temperature profiles, beginning with a single-species ($T_e$-only) evolution and progressing to a coupled two-species ($T_e$ and $T_i$) model including neoclassical transport. Section~\ref{sec:mhd_model} introduces reduced gyrokinetic-based surrogate models for KBM and MHD-like transport constructed from a database of linear gyrokinetic simulations from \textsc{gene}, using equilibria reconstructed self-consistently with the \textsc{spider}~\cite{Ivanov2005} and \textsc{chease}~\cite{Lutjens1996} codes. This section examines the resulting profile predictions, the complementary roles of the transport mechanisms, the role of $E \times B$ shear suppression, and the combined impact of the reduced transport channels. Finally, the main conclusions and perspectives for future model development are summarized in Section~\ref{sec:conclusion}.

\section{Preliminary Profile Predictions with ETG and Neoclassical Transport}
\label{sec:etg_neo}
The reduced ETG transport model employed in this study follows the algebraic form introduced in Ref.~\cite{Hatch2024}:
\begin{equation}
\frac{Q_e}{Q_{GB}} = a_0 \sqrt{\frac{m_e}{m_i}}\, \omega_{Te}^2 (\eta_e - 1)\,\eta_e^{\,b_0}\, \tau^{c_0},
\label{eq:ETGmodel}
\end{equation}
where $Q_e$ is the electron heat flux, $\psi$ is the normalized poloidal magnetic flux serving as the radial coordinate, $\omega_{Te} = \frac{1}{T_e}\frac{dT_e}{d\psi}$ and $\omega_{ne} = \frac{1}{n_e}\frac{dn_e}{d\psi}$ are the normalized temperature and density gradients (with $n_e$ being the electron density and $T_e$ the electron temperature). Furthermore, $\eta_e = \omega_{Te}/\omega_{ne}$, $\tau = Z_{\mathrm{eff}} T_e/T_i$ (where $Z_{\mathrm{eff}}$ is the effective ion charge and $T_i$ is the ion temperature), and $Q_{GB} = n_e T_e c_s \rho_s^2 / a^2$ is the gyro-Bohm heat flux normalization. Here, $m_e$ and $m_i$ are the electron and main ion masses, $c_s = \sqrt{T_e/m_i}$ is the ion sound speed, $\rho_s$ is the ion sound Larmor radius, and $a$ is the tokamak minor radius. The coefficients $a_0=0.019$, $b_0=1.57$, and $c_0=-0.5$ were previously optimized using a nonlinear gyrokinetic database from conventional aspect-ratio tokamaks, as described in Ref.~\cite{Hatch2024}.  Ref.~\cite{Hatch2024} also found good agreement with several nonlinear simulations based on a MAST discharge, finding relatively good agreement, although these spherical tokamak cases were a vast minority compared to standard aspect ratio.

\begin{figure}[ht]
    \centering
    \IfFileExists{NSTX_QMvsQNL.pdf}{\includegraphics[width=0.6\linewidth]{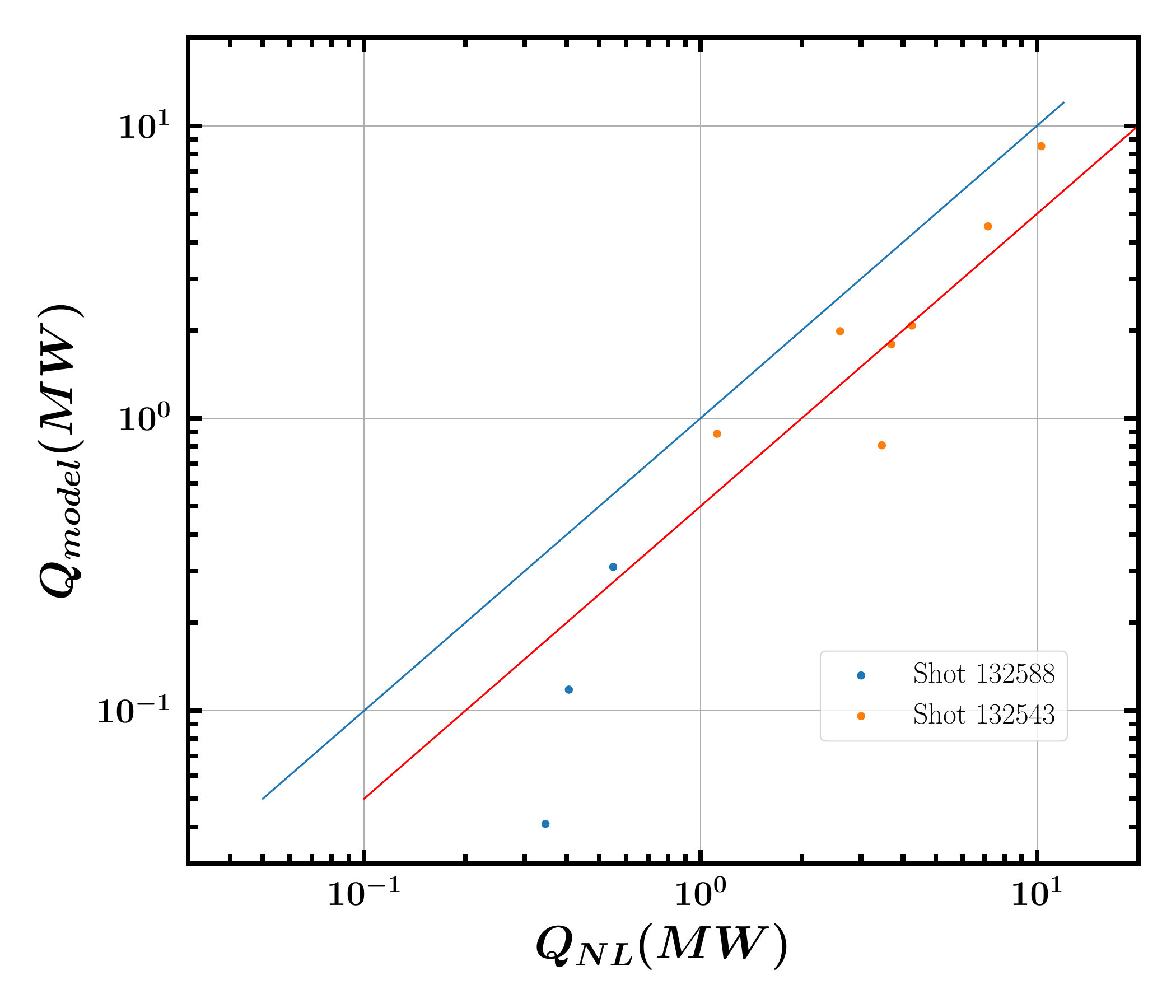}}{\rule{0.6\linewidth}{6cm}}
    \caption{Scatter plot comparing the heat flux from the simple reduced ETG model (y-axis) with numerous nonlinear gyrokinetic simulations from GENE and CGYRO (x-axis) for NSTX. The red line demonstrates that multiplying the calculated heat fluxes from the simple model by a factor of two ($2 \times Q_{\mathrm{model}}$) provides an improved fit to the nonlinear simulations for NSTX. Because the model is stiff with respect to the temperature gradient, these magnitude differences in heat flux result in very small differences for the overall profile prediction. (Simulations performed with the \textsc{gene} code).}
    \label{fig:QMVSQNL}
\end{figure}

To verify its accuracy for the present scenarios, this ETG formula was compared directly with nonlinear gyrokinetic simulations performed using the \textsc{gene} and \textsc{cgyro} codes for NSTX shots 132543 and 132588 at several radial locations. Figure~\ref{fig:QMVSQNL} presents a scatter plot of the heat fluxes calculated from the simple algebraic model against the nonlinear simulation results. As highlighted by the red line in the figure, the reduced ETG formula systematically underestimates the heat flux for these specific NSTX conditions by approximately a factor of two; scaling the model by $2 \times$ yields a better fit to the nonlinear data. 

However, throughout most of this study, we deliberately retain the original, unscaled formula as our baseline. This choice is motivated by the desire for generality; the unscaled algebraic model has been shown to successfully fit a wide range of simulations spanning widely varying scenarios in several conventional tokamaks~\cite{Hatch2024}. Rather than explicitly tuning the model for NSTX geometry at the outset, we rely on the formula's strong sensitivity to the temperature gradient (scaling approximately as $\omega_{Te}^{4.5}$). Because this transport is stiff, even a small increase in the gradient produces a large rise in heat flux, rapidly restoring the transport balance without requiring manual coefficient calibration. We will explicitly evaluate the profile impact of the $2 \times Q_{\mathrm{ETG}}$ scaling later in this section.

The profile predictions in this study are performed using the Automated System for TRansport Analysis (\textsc{astra}) code~\cite{Pereverzev2002}. \textsc{astra} is a comprehensive 1.5D transport solver that calculates the time-dependent evolution of plasma profiles (such as temperatures and densities) by solving a set of coupled radial transport equations. To accurately capture the specific experimental equilibrium of the NSTX discharges, \textsc{astra} was initialized with several key macroscopic parameters: the shape of the last closed magnetic flux surface, the toroidal magnetic field on axis, the total plasma current, and the major radius. Neutral beam heating power and particle sources were prescribed based on experimental data. Using these boundary inputs alongside the reconstructed Grad-Shafranov equilibria, \textsc{astra} maps the 2D magnetic geometry and calculates the necessary flux-surface-averaged metric coefficients. 

In \textsc{astra}, the temperature evolution is calculated by determining the thermal diffusivities ($\chi$) rather than directly prescribing heat fluxes. In order to calculate the pedestal's temperature profile, the total effective thermal diffusivity for each species incorporates both the turbulent ETG formulation and standard neoclassical components computed self-consistently utilizing the NCLASS module within \textsc{astra}:
\begin{align}
\chi_e &= \chi_{e,\mathrm{ETG}} + \chi_{e,\mathrm{neo}}, \label{eq:chie}\\
\chi_i &= \chi_{i,\mathrm{neo}}, \label{eq:chii}
\end{align}
where the turbulent ETG diffusivity is derived directly from the modeled heat flux via $\chi_{e,\mathrm{ETG}} = Q_{e,\mathrm{ETG}} / (n_e |\nabla T_e|)$ using Eq.~\ref{eq:ETGmodel}. Furthermore, the dynamic interplay between the evolution of the electron and ion thermal channels is mediated by the electron-ion collisional power transfer, $P_{ei}$. Depending on the local temperature differential, $P_{ei}$ acts as a coupled volumetric source and sink within the respective energy transport equations. Within this framework, \textsc{astra} evaluates the local effective thermal diffusivities at each time step and dynamically evolves the temperature profiles toward a steady state. Section~\ref{sec:ETG_only} will first isolate the turbulent electron channel by evaluating a simplified subset of these equations.

\subsection{Single-species ETG-only evolution with fixed density and ion temperature}
\label{sec:ETG_only}

Before evaluating the simulated transport, it is important to establish the contrasting experimental scenarios of the two representative high triangularity NSTX discharges analyzed in this study. For discharge 132588, which features a lithiated wall and a relatively wide pedestal, the experimental data and the corresponding profile used for analysis are at time 650~ms. Because this discharge is ELM-free, the temperature profiles have had sufficient time to fully saturate, meaning the pedestal operates in a steady state. In contrast, for discharge 132543, a non-lithiated, ELMy plasma characterized by a narrower pedestal, the experimental data and the corresponding profile used for analysis are at time 700~ms, which falls during a post-ELM recovery phase. Because these measurements were taken in the inter-ELM period, the experimental temperature profiles are still slowly growing and are not in a steady state. Consequently, when applying a steady-state transport solver like \textsc{astra} to discharge 132543, we inherently expect the predicted temperatures to slightly overpredict the transient experimental measurements.

With these operational contexts established, an initial modeling phase was performed to isolate the influence of electron temperature gradient (ETG) transport on the pedestal temperature profiles. In this phase, only a single species---the electron temperature profile---was allowed to evolve ($\chi_e = \chi_{e,\mathrm{ETG}}$ with neoclassical terms disabled). The experimental density profile $n_e(\rho_{tor})$ and ion temperature profile $T_i(\rho_{tor})$ (where $\rho_{tor}$ is the normalized toroidal magnetic flux coordinate) were held fixed, serving as a controlled baseline to evaluate the predictive capability of the reduced formula. 

\begin{figure}[ht]
    \centering
    \begin{subfigure}{0.48\textwidth}
        \IfFileExists{132543_e_T.pdf}{\includegraphics[width=\linewidth]{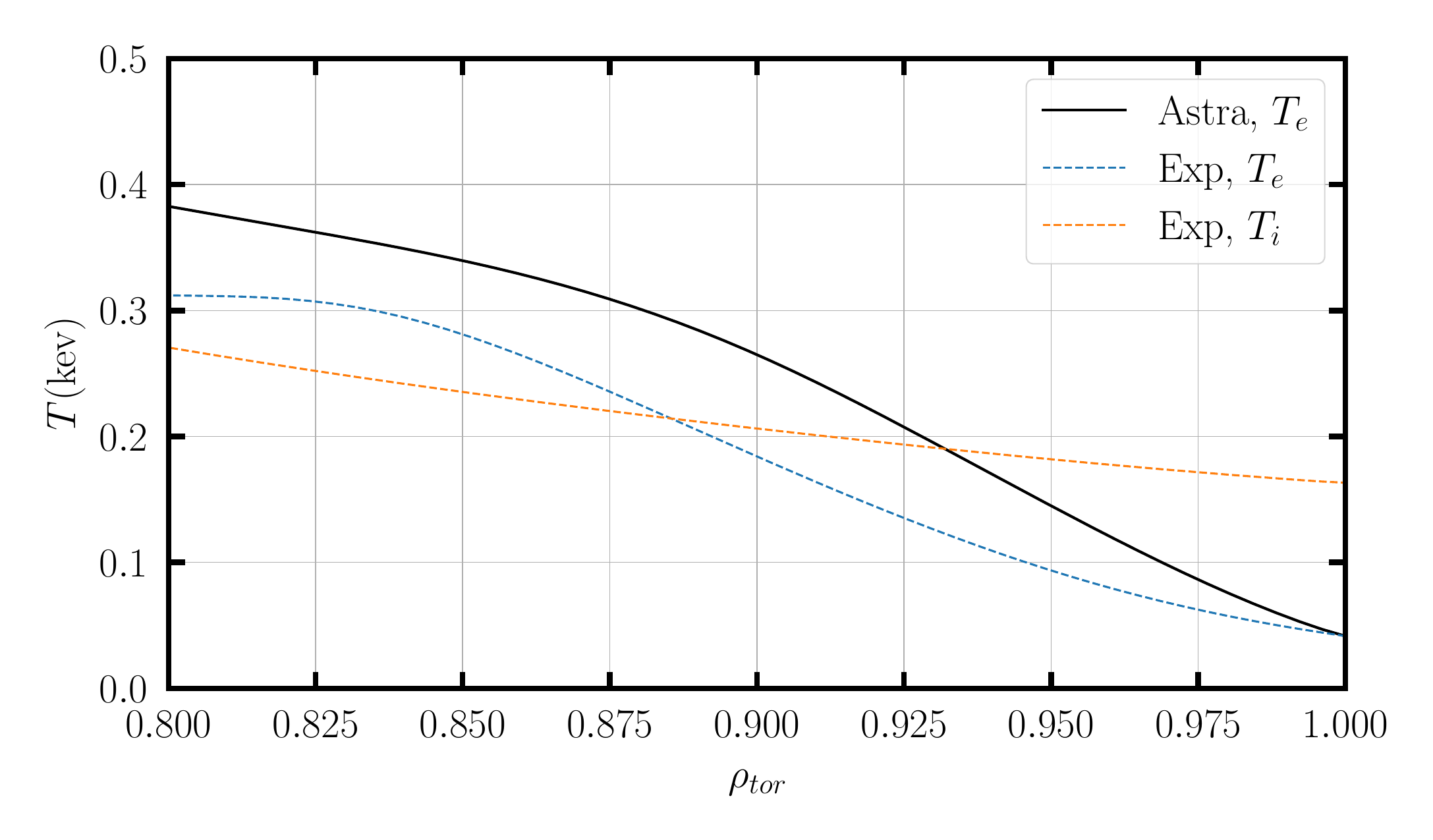}}{\rule{\linewidth}{6cm}}
        \caption{132543}
    \end{subfigure}
    \hfill
    \begin{subfigure}{0.48\textwidth}
        \IfFileExists{132588_e_T.pdf}{\includegraphics[width=\linewidth]{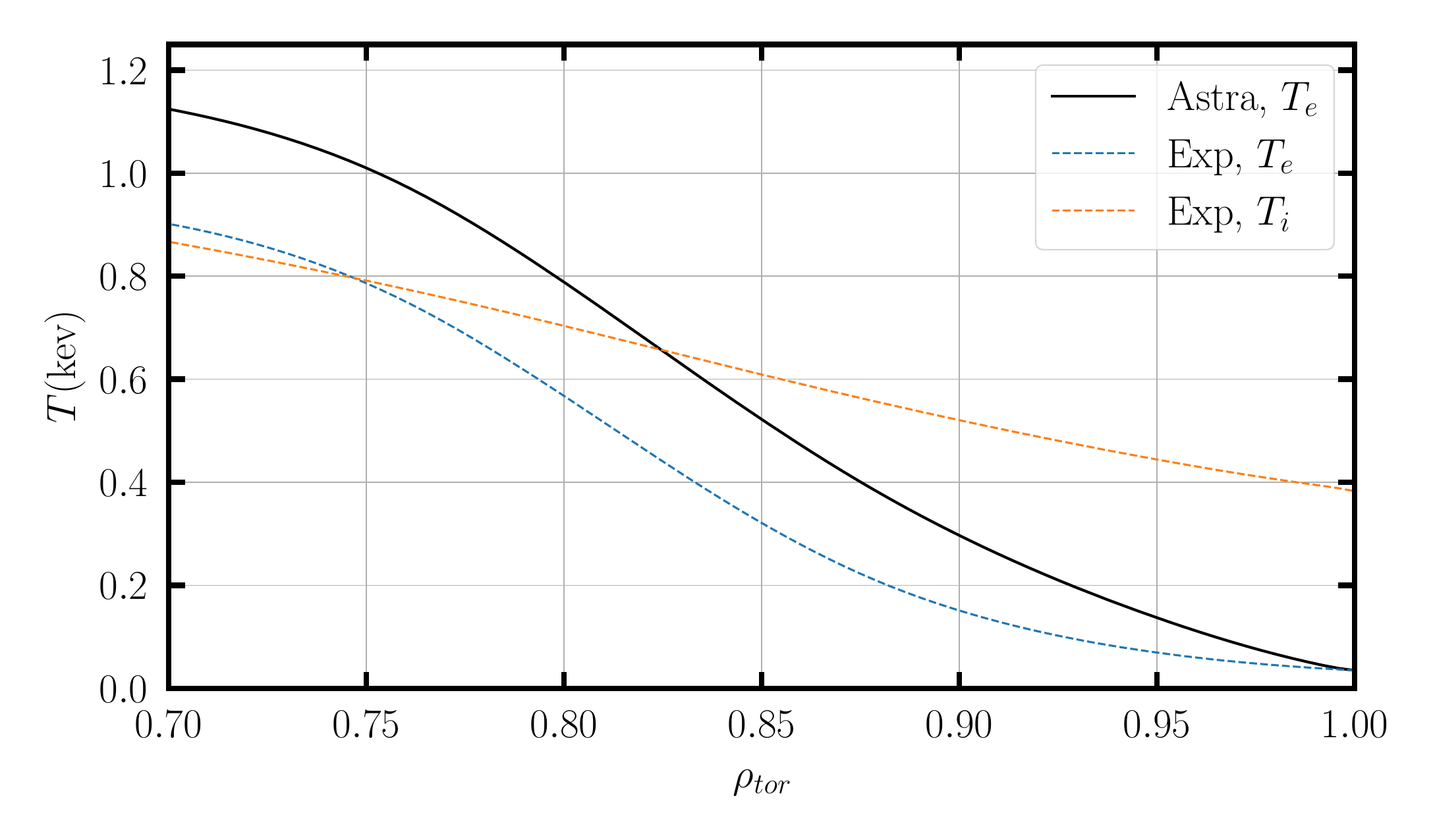}}{\rule{\linewidth}{6cm}}
        \caption{132588}
    \end{subfigure}
    \caption{$T_e$ profile prediction from the unscaled baseline model using ASTRA with fixed $T_i$ and $n_e$. The predicted profiles find reasonable agreement with experimental measurements, though some minor differences remain for (a) shot 132543 and (b) shot 132588.}
    \label{fig:etg_only_profile}
\end{figure}

\begin{figure}[ht]
    \centering
    \begin{subfigure}{0.48\textwidth}
        \IfFileExists{132543_e_chi.pdf}{\includegraphics[width=\linewidth]{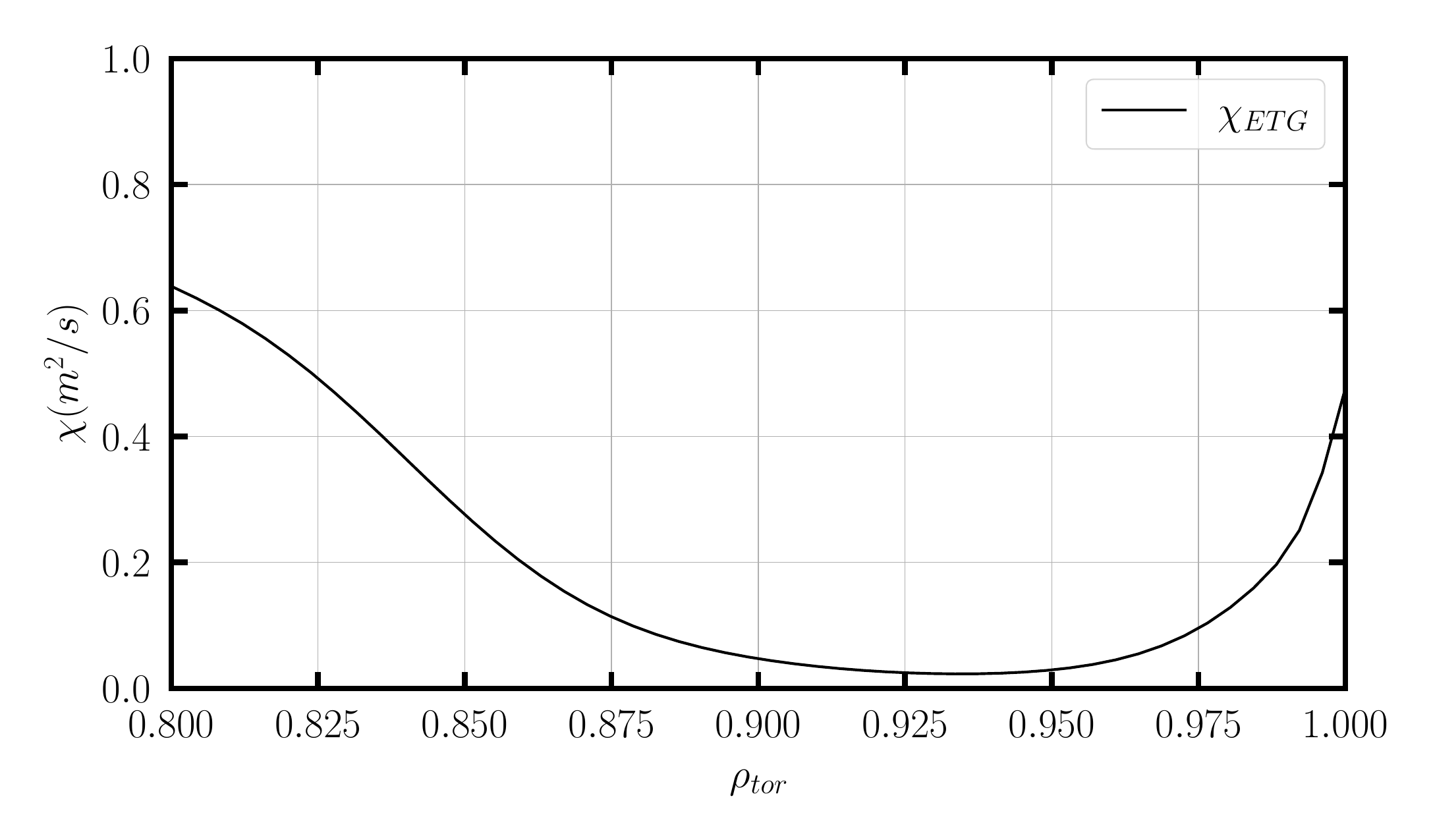}}{\rule{\linewidth}{6cm}}
        \caption{132543}
    \end{subfigure}
    \hfill
    \begin{subfigure}{0.48\textwidth}
        \IfFileExists{132588_e_chi.pdf}{\includegraphics[width=\linewidth]{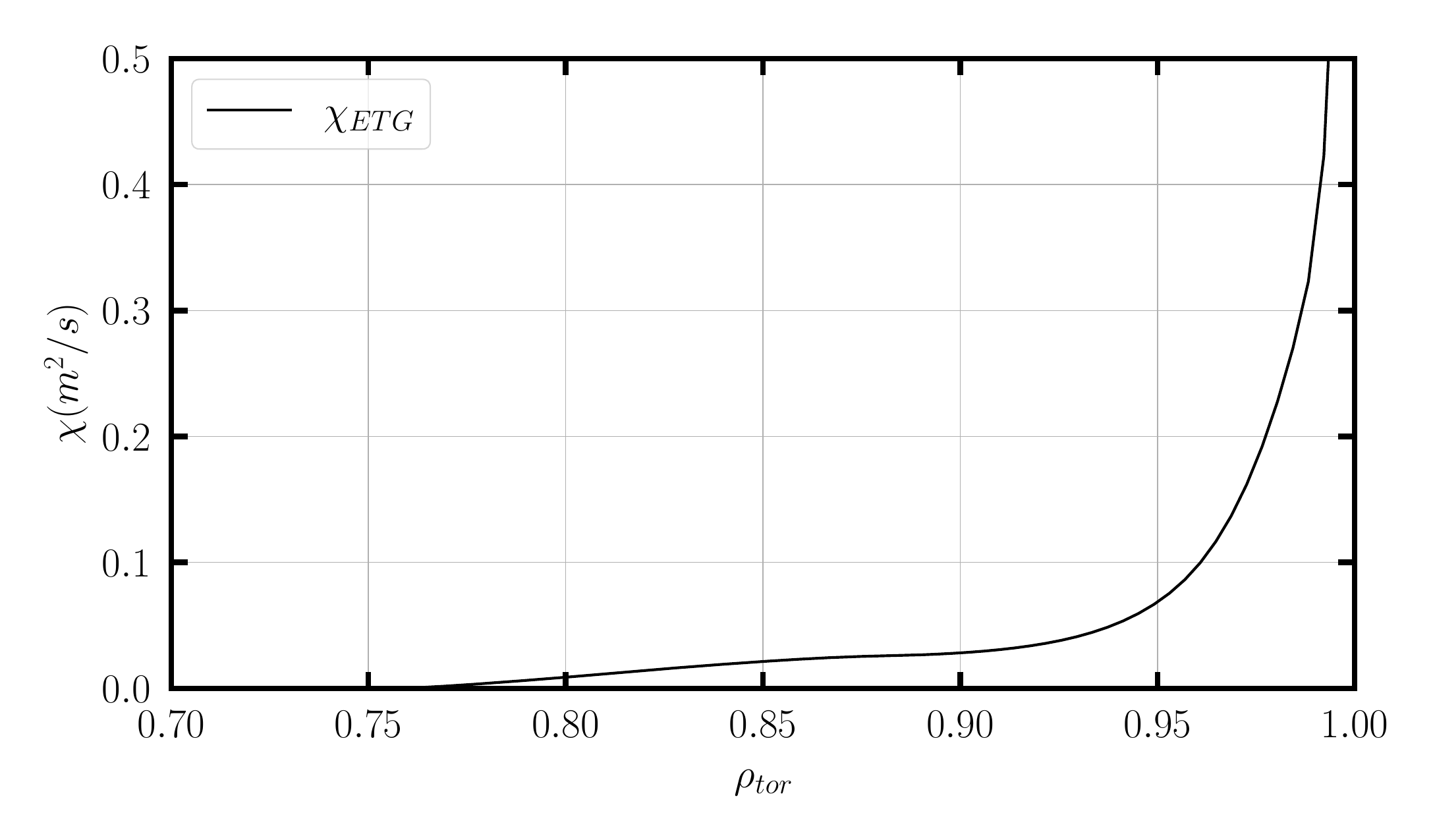}}{\rule{\linewidth}{6cm}}
        \caption{132588}
    \end{subfigure}
    \caption{Thermal diffusivity ($\chi_e$) profiles driven by ETG for (a) 132543 and (b) 132588. The results show that the ETG-driven transport is particularly strong and highly localized in the steep gradient region of the pedestal.}
    \label{fig:qetg_profile}
\end{figure}

\begin{figure}[ht]
    \centering
    \begin{subfigure}{0.48\textwidth}
        \IfFileExists{132543_e_Pei.pdf}{\includegraphics[width=\linewidth]{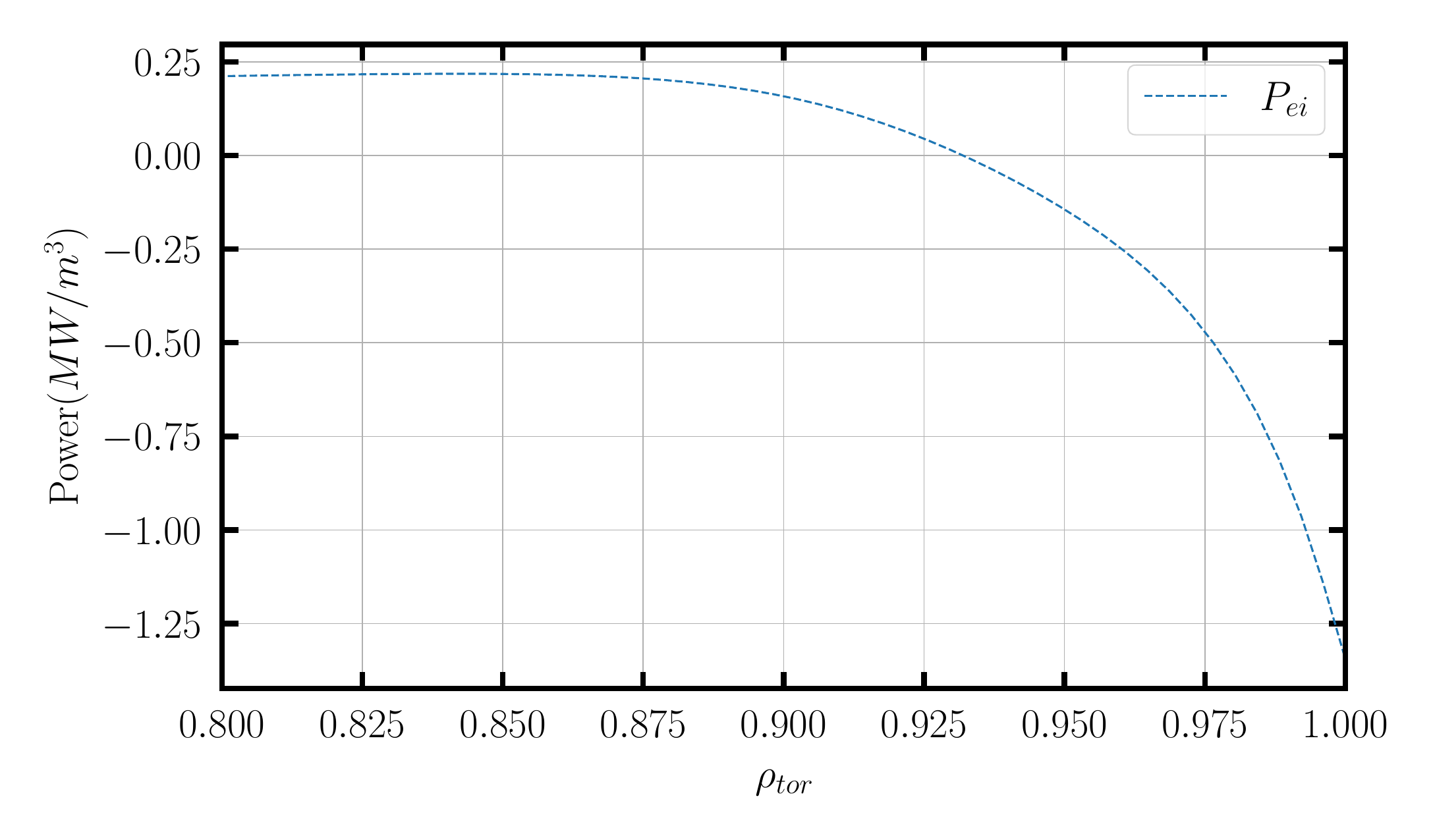}}{\rule{\linewidth}{6cm}}
        \caption{132543}
    \end{subfigure}
    \hfill
    \begin{subfigure}{0.48\textwidth}
        \IfFileExists{132588_e_Pei.pdf}{\includegraphics[width=\linewidth]{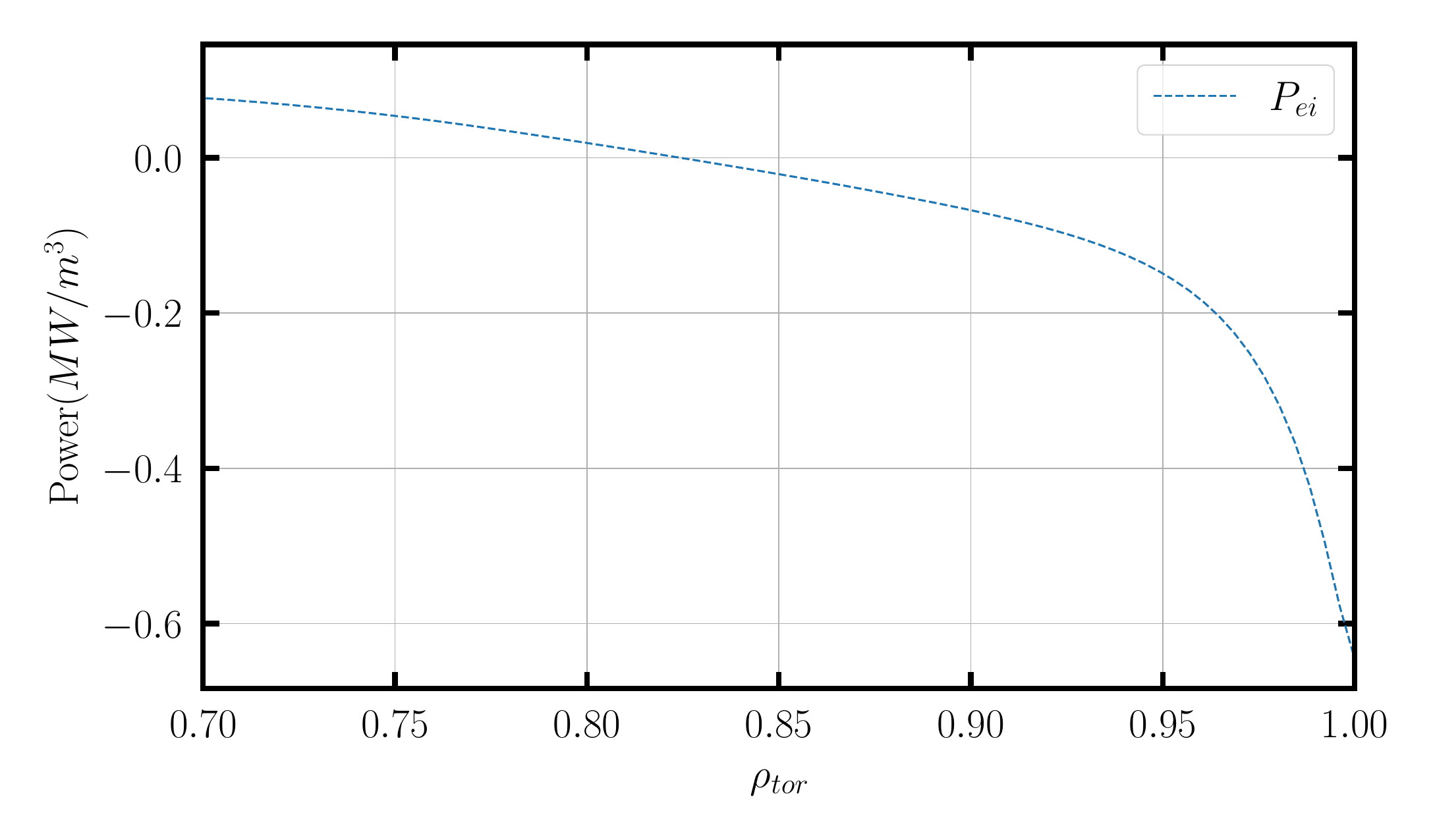}}{\rule{\linewidth}{6cm}}
        \caption{132588}
    \end{subfigure}
    \caption{The electron-ion energy exchange rate per unit volume ($P_{ei}$) for (a) 132543 and (b) 132588. Note the sign convention where $P_{ei}$ represents energy transfer into the ions from the electrons. $P_{ei} < 0$ indicates energy is transferred to the electrons in the steep gradient region, while $P_{ei} > 0$ indicates energy is transferred to the ions in the regions with a smaller $T_e$ gradient.}
    \label{fig:pei_profile}
\end{figure}

Figure~\ref{fig:etg_only_profile} demonstrates that this setup yields reasonable steady-state $T_e$ profile predictions that capture the qualitative shape of the experimental data. Examining the underlying physics driving these profiles reveals a highly localized spatial transport distribution. As shown in Figure~\ref{fig:qetg_profile}, the ETG-driven thermal diffusivity ($\chi_e$) peaks sharply in the steep gradient region of the pedestal. This is consistent with expectations: ETG modes are highly unstable in this narrow region due to the large values of $\omega_{Te}$ combined with local conditions where $\eta_e \gg 1$ (i.e., where the density gradient is small relative to the temperature gradient).

Coupled with this turbulent transport is the classical electron-ion collisional energy exchange, $P_{ei}$, shown in Figure~\ref{fig:pei_profile}. We define $P_{ei}$ as the energy transfer into the ions from the electrons (i.e., $P_{ei} > 0$ implies energy flowing from electrons to ions). The energy exchange rate exhibits a distinct spatial dipole. In the steep gradient region where ETG turbulence is strongest, $P_{ei}$ is negative, meaning energy flows from the ions to the electrons, acting as a powerful local heating source for the electron channel. Conversely, further radially inward where the $T_e$ gradient relaxes, the temperature differential changes and the flow reverses ($P_{ei} > 0$), transferring energy from the electrons back to the ions. 

While this baseline ETG-only configuration successfully reproduces the overall shape of the electron temperature profile, the modeled electron temperature is still somewhat overpredicted compared to the experiment. This overprediction indicates that the transport provided by the unscaled ETG model is slightly insufficient to fully relax the pedestal. Therefore, while ETG turbulence captures the vast majority of the electron heat transport, additional electron transport mechanisms---such as microtearing modes (MTM)~\cite{Guttenfelder2012}---should be considered in future work to achieve a better matched theoretical result. However, as demonstrated later in the fully coupled models, MTM is likely a secondary effect after ETG and KBM, since these two mechanisms combined can capture most of the relevant transport effects already.

\begin{figure}[ht]
    \centering
    \begin{subfigure}{0.48\textwidth}
        \IfFileExists{132543_e_T_nscan.pdf}{\includegraphics[width=\linewidth]{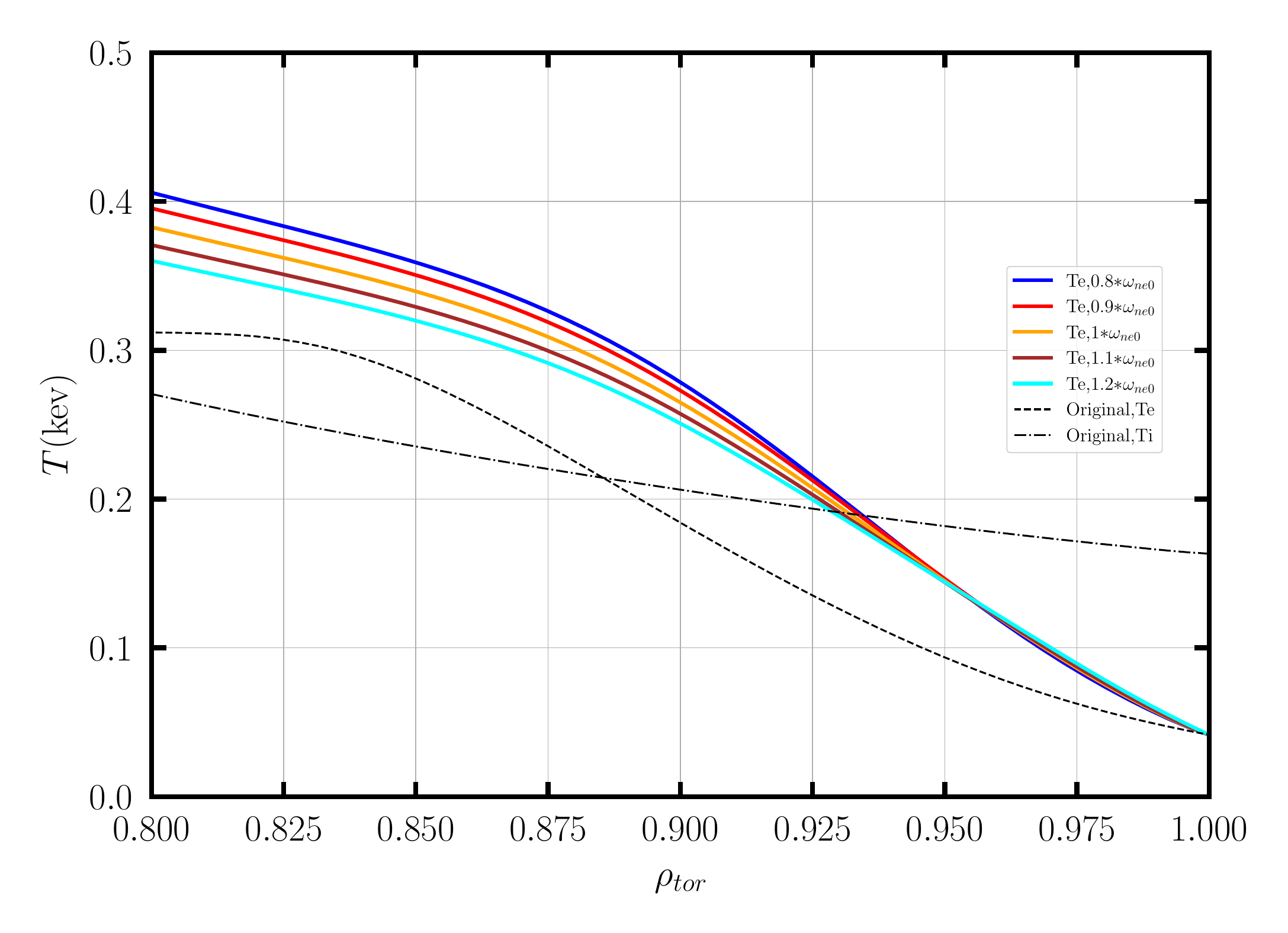}}{\rule{\linewidth}{6cm}}
        \caption{132543, pivot $\rho=1.0$}
    \end{subfigure}
    \hfill
    \begin{subfigure}{0.48\textwidth}
        \IfFileExists{132588_e_T_nscan.pdf}{\includegraphics[width=\linewidth]{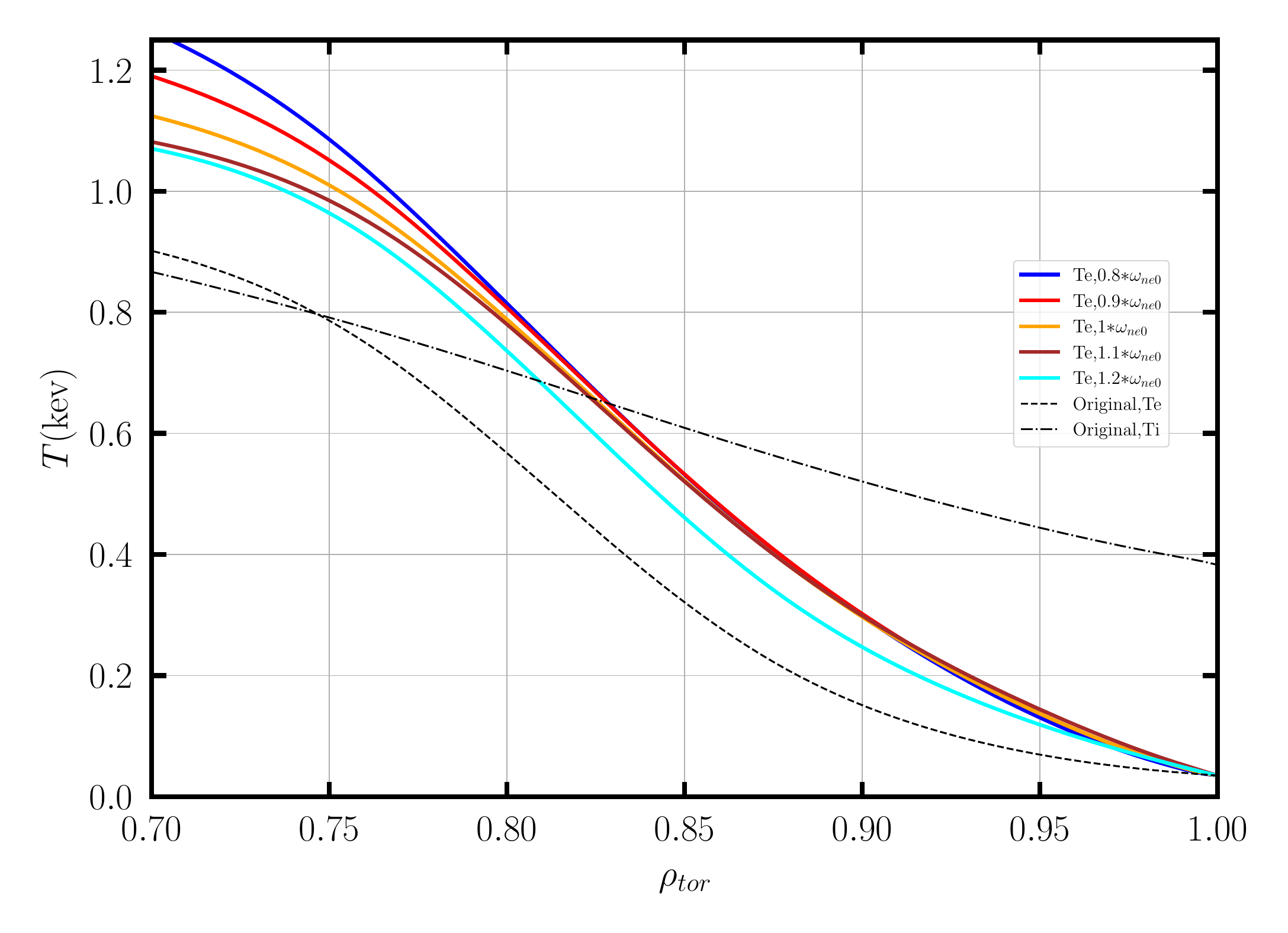}}{\rule{\linewidth}{6cm}}
        \caption{132588, pivot $\rho=1.0$}
    \end{subfigure}\\
    \vspace{0.5cm}
    \begin{subfigure}{0.48\textwidth}
        \IfFileExists{132543_e_T_nscan_nfix09.pdf}{\includegraphics[width=\linewidth]{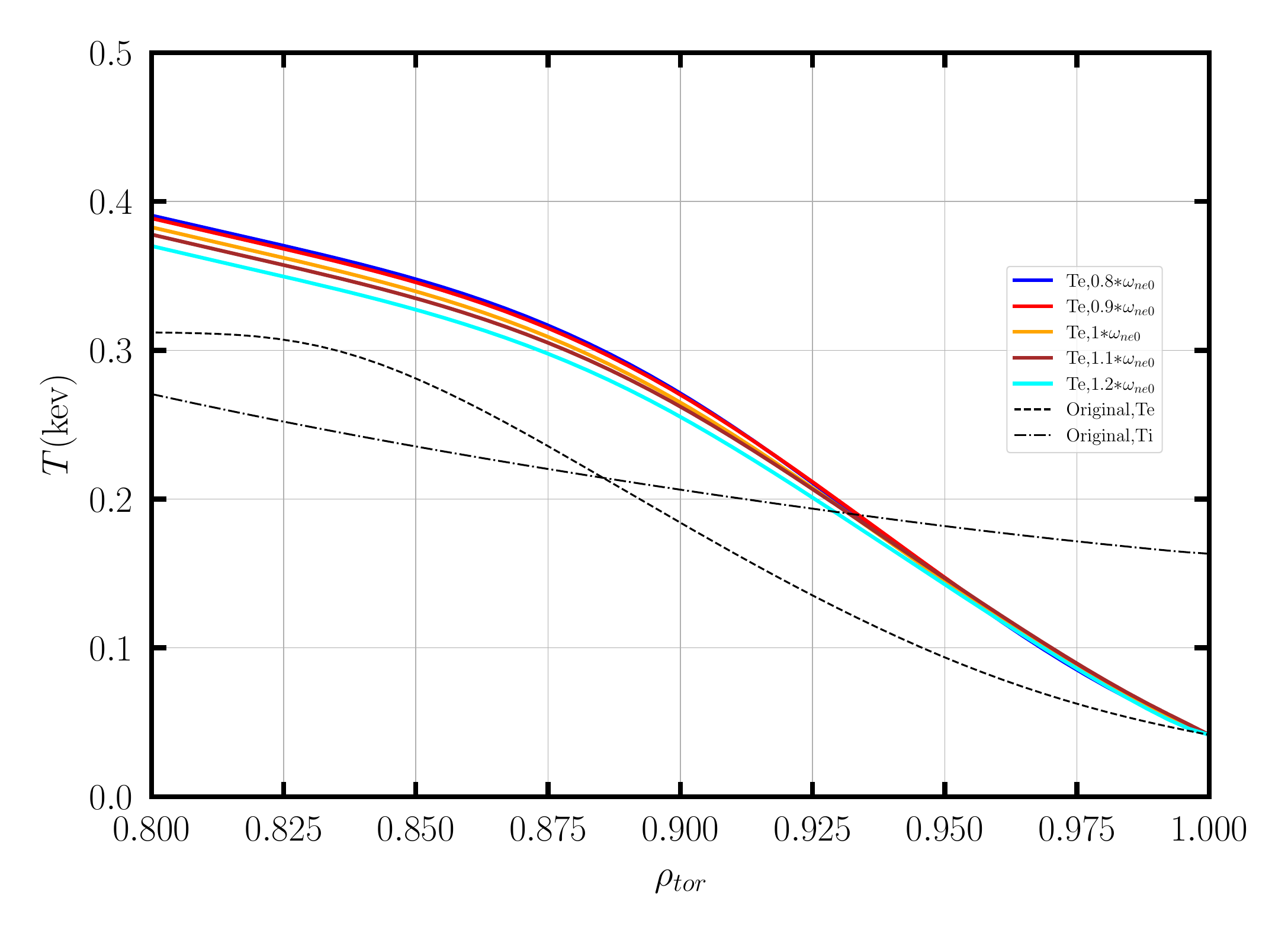}}{\rule{\linewidth}{6cm}}
        \caption{132543, pivot $\rho=0.9$}
    \end{subfigure}
    \hfill
    \begin{subfigure}{0.48\textwidth}
        \IfFileExists{132588_e_T_nscan_nfix09.pdf}{\includegraphics[width=\linewidth]{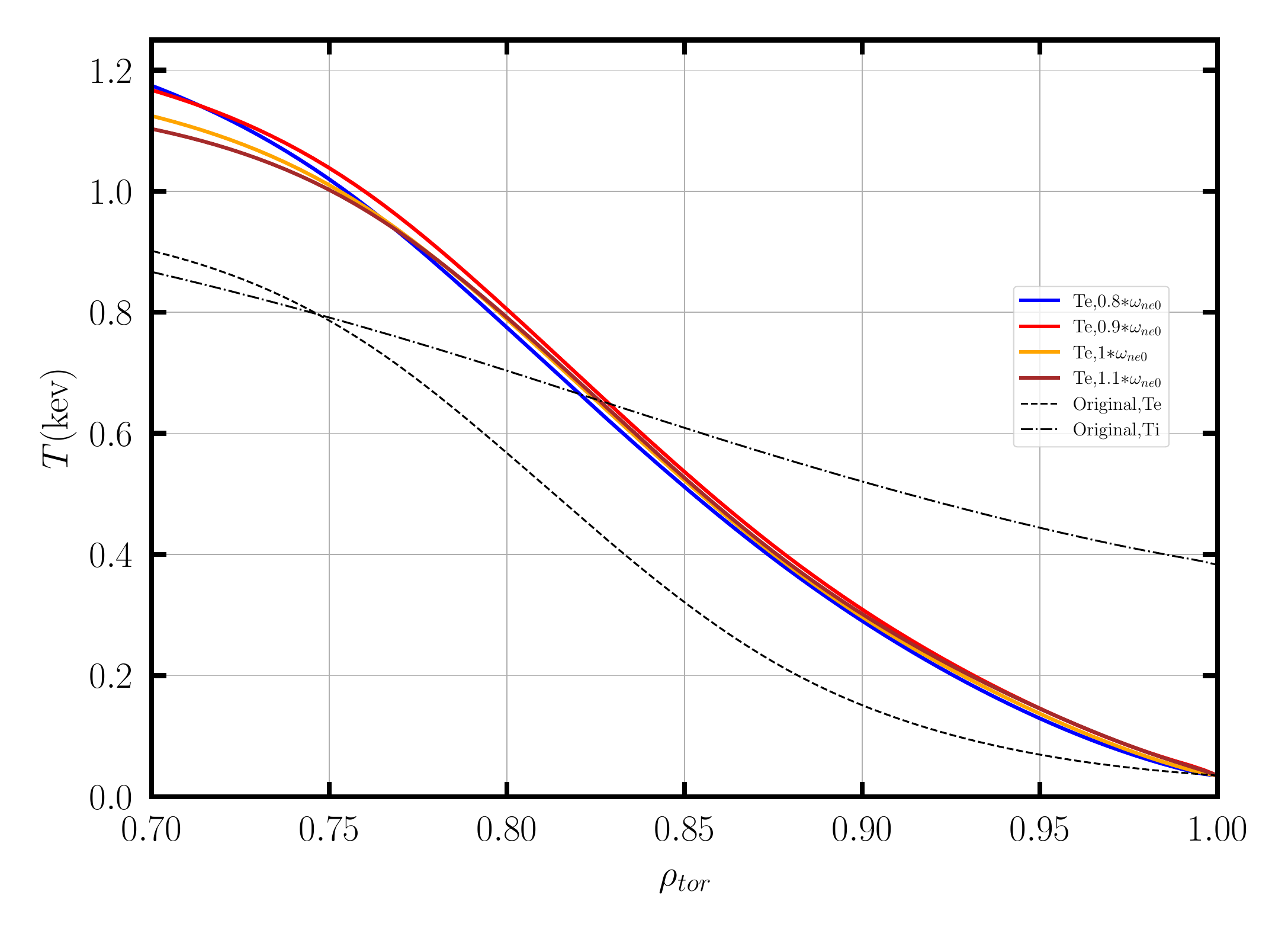}}{\rule{\linewidth}{6cm}}
        \caption{132588, pivot $\rho=0.9$}
    \end{subfigure}
    \caption{$T_e$ profiles predicted with different density profiles (increasing or decreasing the density gradients by $\pm 20\%$). Panels (a) 132543 and (b) 132588 use a pivot point for the density profile adjustment at $\rho=1.0$ from the original one. Panels (c) 132543 and (d) 132588 use a pivot point at $\rho=0.9$. Overall, the $T_e$ profile is highly resilient and does not change much.}
    \label{fig:te_dens_scan}
\end{figure}

Given the strong dependence of ETG turbulence on the parameter $\eta_e = L_n / L_{Te}$ (where $L_n$ and $L_{Te}$ are the respective density and electron temperature gradient scale lengths), it is critical to evaluate how sensitive the single-species $T_e$ predictions are to variations in the prescribed density profile. We artificially steepened and flattened the density gradients by $\pm 20\%$ at two different radial pivot points ($\rho=1.0$ and $\rho=0.9$). 

As shown in Figure~\ref{fig:te_dens_scan}, the predicted electron temperature profile is remarkably resilient to these density modifications. The physical mechanism ensuring this robustness is governed by a $P_{ei}$ cancellation effect. When the density gradient is flattened (larger $L_n$), the value of $\eta_e$ increases, which intrinsically raises the strength of the ETG turbulent drive and should theoretically cool the electrons. However, the classical energy exchange $P_{ei}$ scales heavily with density. A lower density profile weakens the collisional coupling, thereby decreasing the energy transferred from the hotter ions to the electrons in the steep gradient region. These two effects---an increase in turbulent thermal transport outwards and a relative decrease in the local electron collisional heating source---effectively cancel each other out. This dynamic buffering locks the $T_e$ profile in place despite significant alterations to the underlying density profile.

\begin{figure}[ht]
    \centering
    \begin{subfigure}{0.48\textwidth}
        \IfFileExists{132543_e_T2.pdf}{\includegraphics[width=\linewidth]{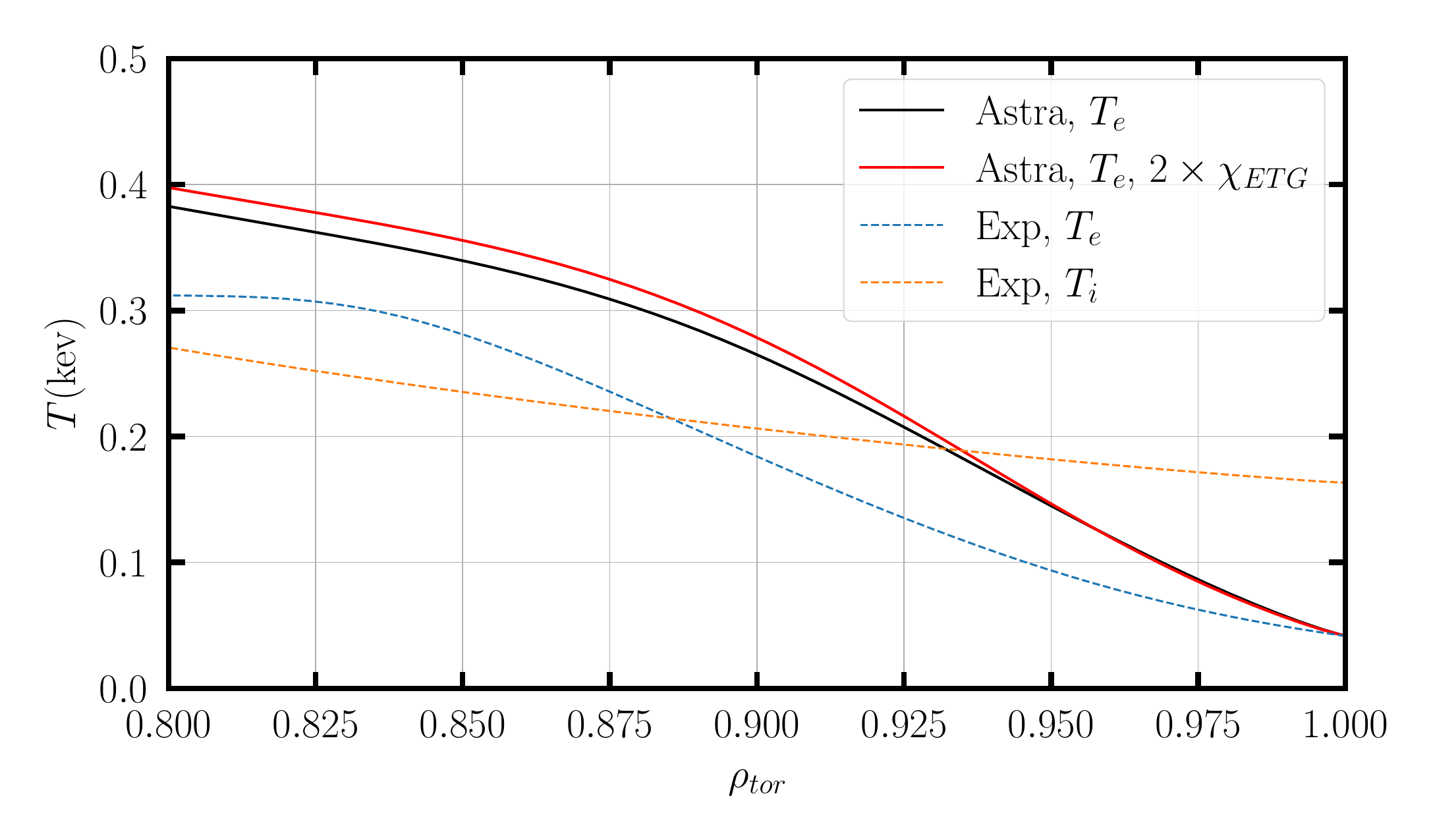}}{\rule{\linewidth}{6cm}}
        \caption{132543}
    \end{subfigure}
    \hfill
    \begin{subfigure}{0.48\textwidth}
        \IfFileExists{132588_e_T2.pdf}{\includegraphics[width=\linewidth]{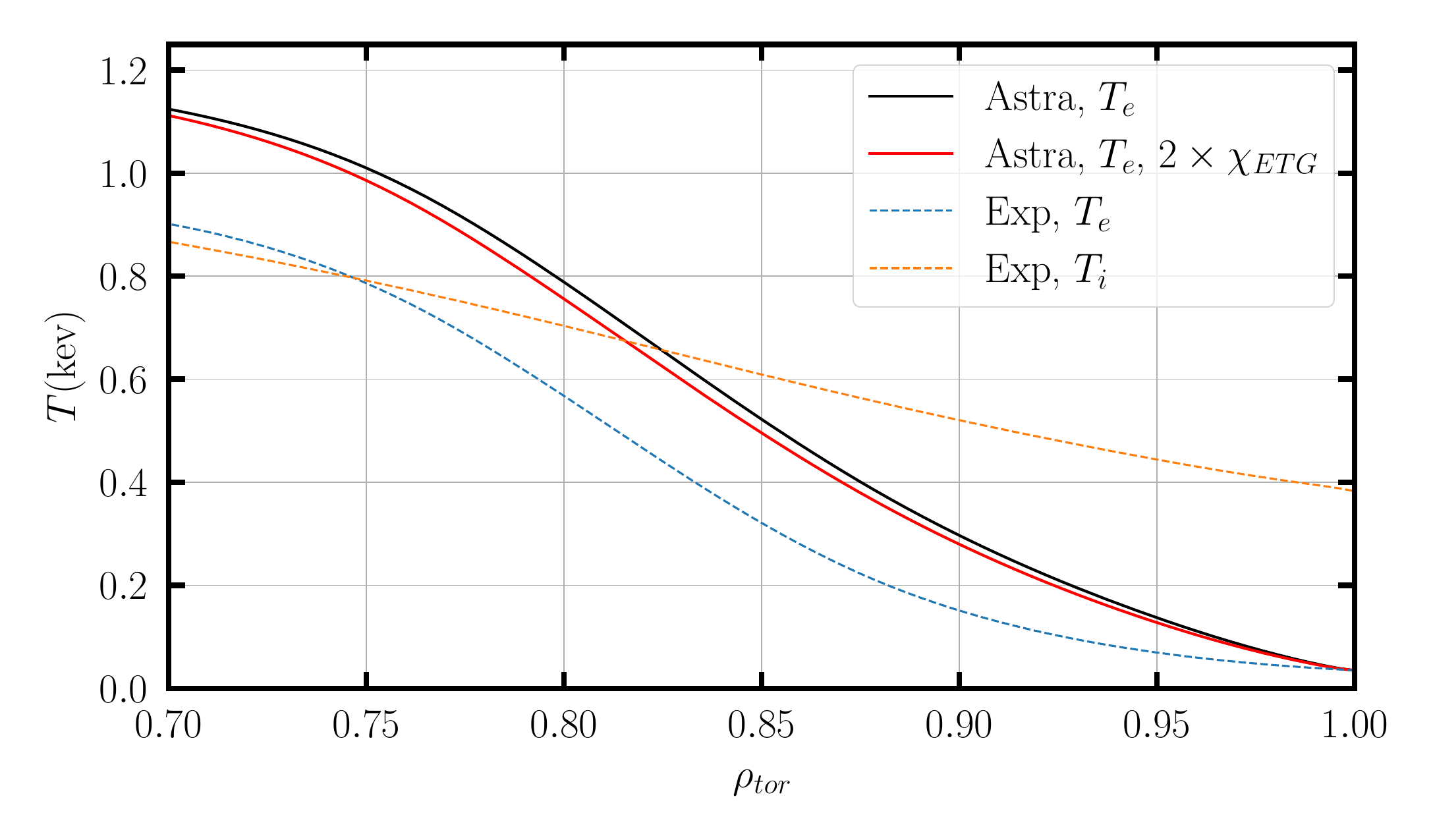}}{\rule{\linewidth}{6cm}}
        \caption{132588}
    \end{subfigure}
    \caption{Single-species $T_e$ profile prediction using ASTRA with fixed $T_i$ and $n_e$, incorporating the scaled $2 \times Q_{\mathrm{ETG}}$ transport. Despite doubling the transport coefficient, the overall magnitude of the $T_e$ profile barely changes from the baseline prediction due to strong $P_{ei}$ cancellation.}
    \label{fig:single_spec_2x}
\end{figure}

To explicitly evaluate the impact of the scaling factor identified in Figure~\ref{fig:QMVSQNL}, we performed an additional single-species simulation where the ETG-driven thermal diffusivity was multiplied by a factor of two ($\chi_e = 2 \times \chi_{e,\mathrm{ETG}}$). Surprisingly, as shown in Figure~\ref{fig:single_spec_2x}, this modification barely changes the absolute magnitude of the predicted $T_e$ profile. This resilience is a direct consequence of the aforementioned $P_{ei}$ cancellation effect: while the doubled ETG drive increases the outward heat flux, it simultaneously alters the temperature differential between the evolving electrons and the fixed ions, causing a strong compensatory increase in the $P_{ei}$ heating source (from ions to electrons). These effects balance, effectively buffering the $T_e$ magnitude against the increased transport, though the scaling does introduce very slight changes to the profile shape.

\subsection{Two-species ETG and neoclassical coupling with evolving ion and electron temperatures}

\begin{figure}[ht]
    \centering
    \begin{subfigure}{0.48\textwidth}
        \IfFileExists{132543_T.pdf}{\includegraphics[width=\linewidth]{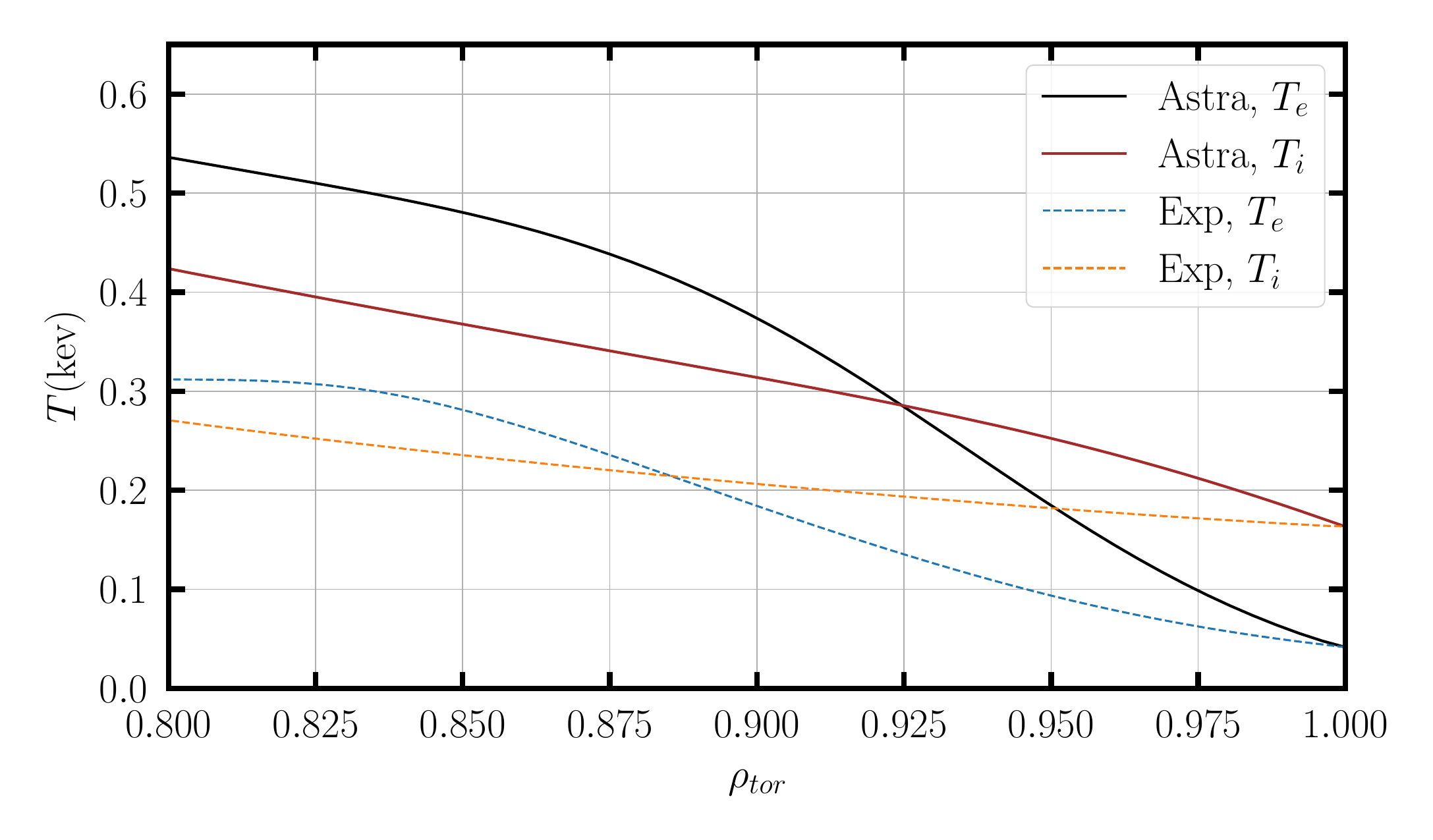}}{\rule{\linewidth}{6cm}}
        \caption{132543}
    \end{subfigure}
    \hfill
    \begin{subfigure}{0.48\textwidth}
        \IfFileExists{132588_T.pdf}{\includegraphics[width=\linewidth]{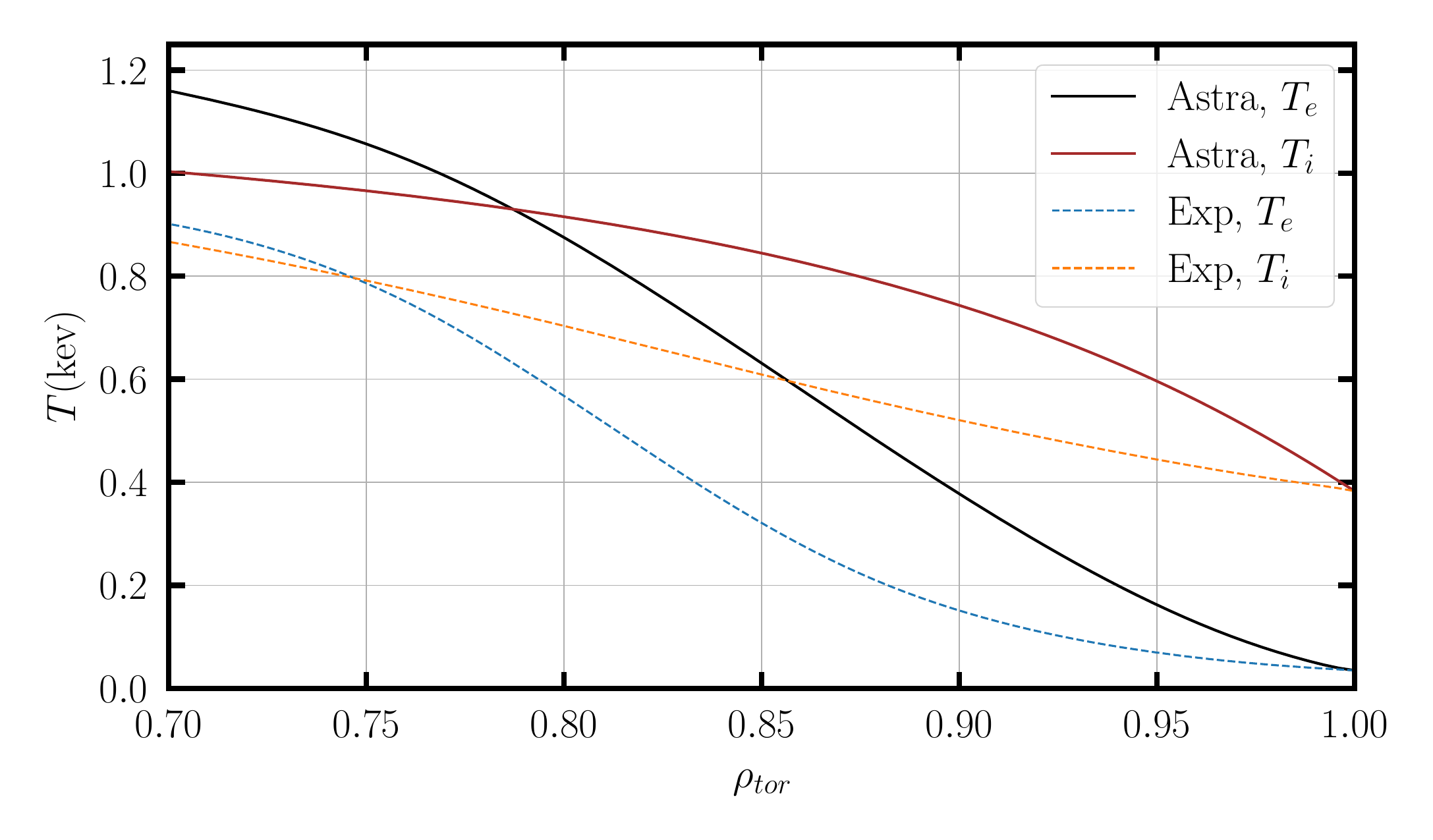}}{\rule{\linewidth}{6cm}}
        \caption{132588}
    \end{subfigure}
    \caption{$T_e$ profile prediction from the unscaled baseline model using ASTRA with fixed $n_e$ (allowing $T_i$ to evolve). The profiles capture the general shape but systematically overpredict the experimental pedestal temperatures for both (a) 132543 and (b) 132588.}
    \label{fig:te_profile_neo}
\end{figure}

\begin{figure}[ht]
    \centering
    \begin{subfigure}{0.48\textwidth}
        \IfFileExists{132543_chi.pdf}{\includegraphics[width=\linewidth]{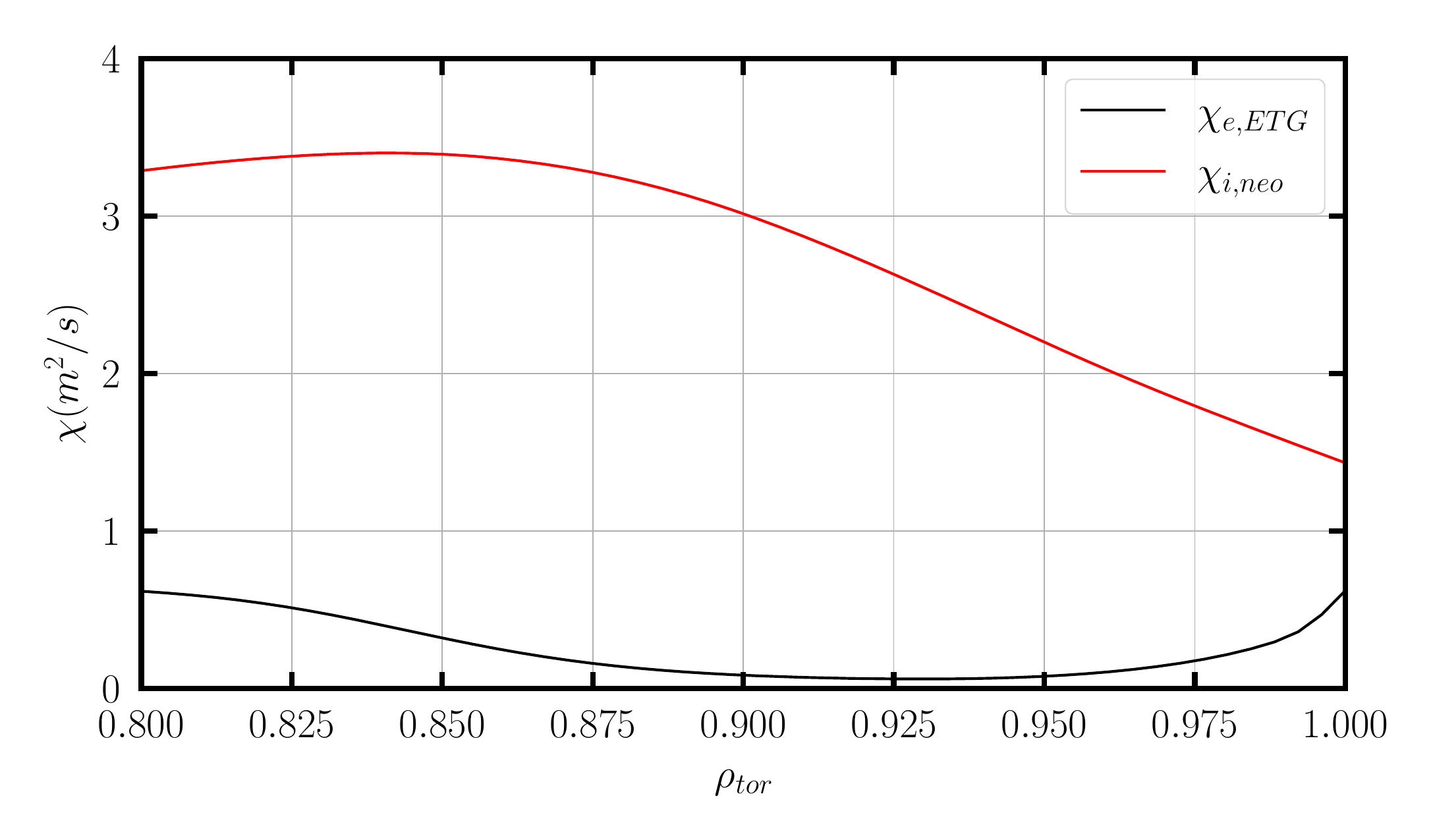}}{\rule{\linewidth}{6cm}}
        \caption{132543}
    \end{subfigure}
    \hfill
    \begin{subfigure}{0.48\textwidth}
        \IfFileExists{132588_chi.pdf}{\includegraphics[width=\linewidth]{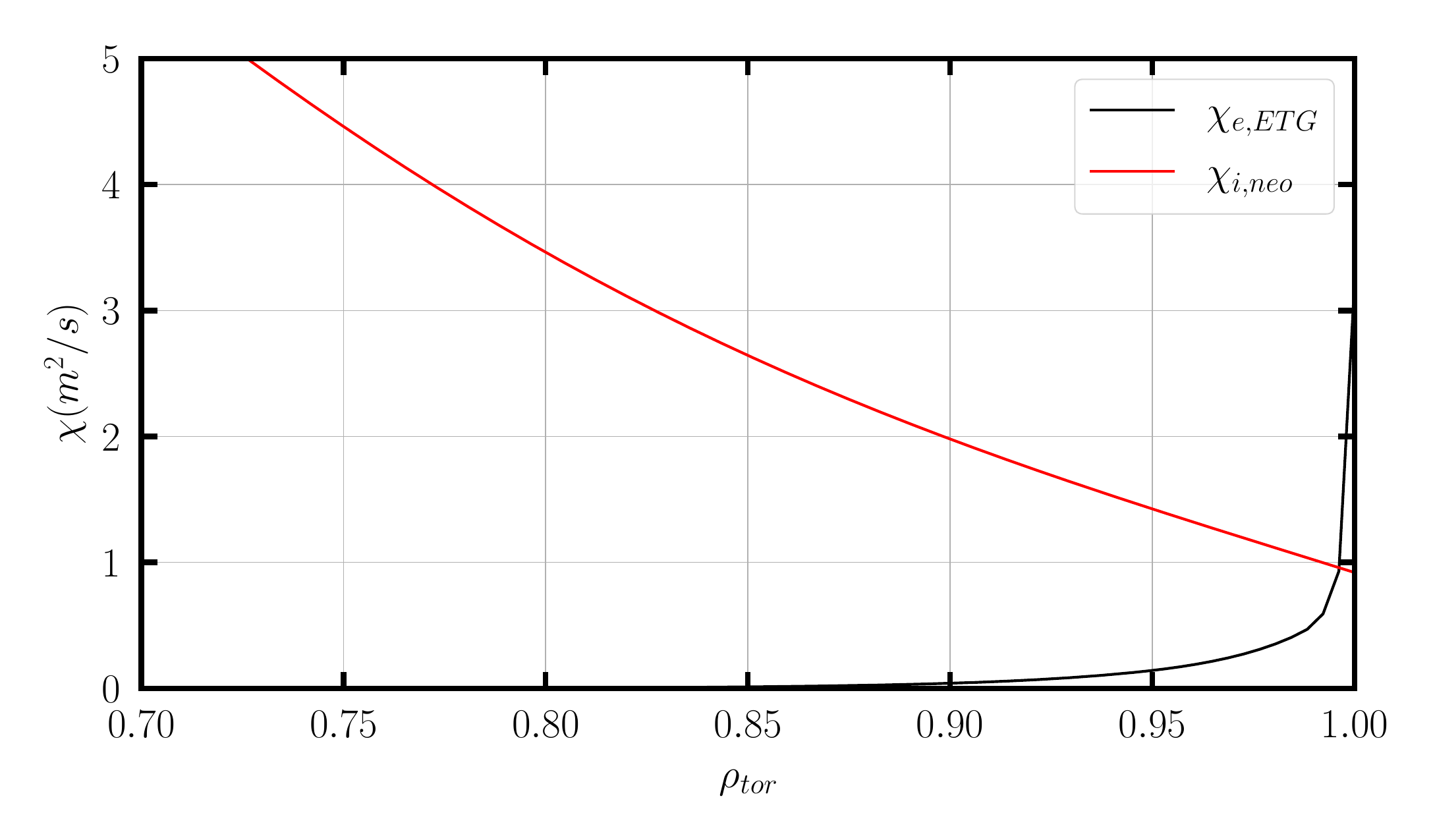}}{\rule{\linewidth}{6cm}}
        \caption{132588}
    \end{subfigure}
    \caption{Thermal diffusivity ($\chi$) profiles separated into ETG and neoclassical components for (a) 132543 and (b) 132588. Neoclassical transport provides a consistently large baseline across the profile, while ETG-driven transport is highly localized to the pedestal top and plasma edge where $\eta_e$ is large.}
    \label{fig:q_profile_neo}
\end{figure}

\begin{figure}[ht]
    \centering
    \begin{subfigure}{0.48\textwidth}
        \IfFileExists{132543_Pei.pdf}{\includegraphics[width=\linewidth]{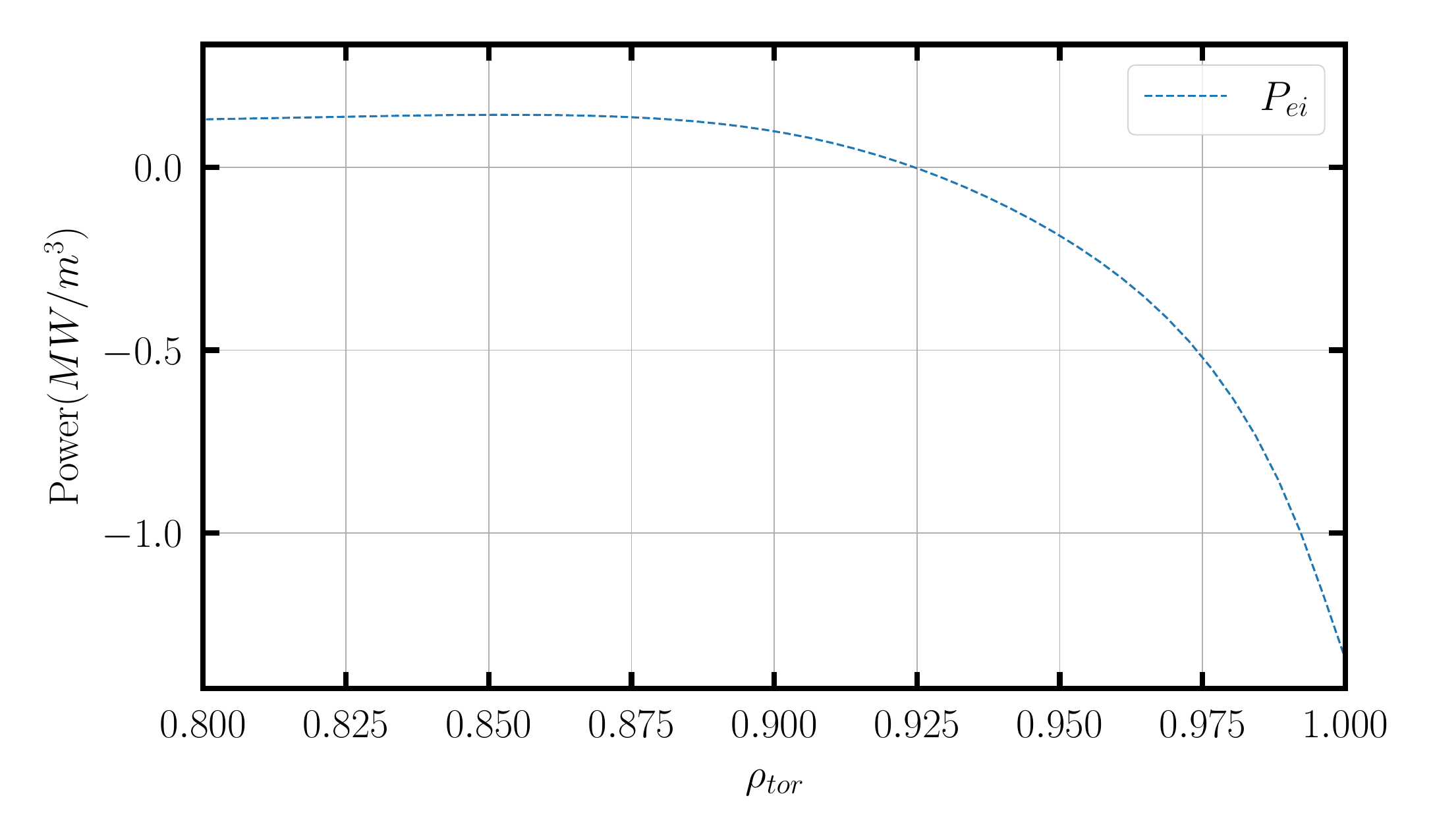}}{\rule{\linewidth}{6cm}}
        \caption{132543}
    \end{subfigure}
    \hfill
    \begin{subfigure}{0.48\textwidth}
        \IfFileExists{132588_Pei.pdf}{\includegraphics[width=\linewidth]{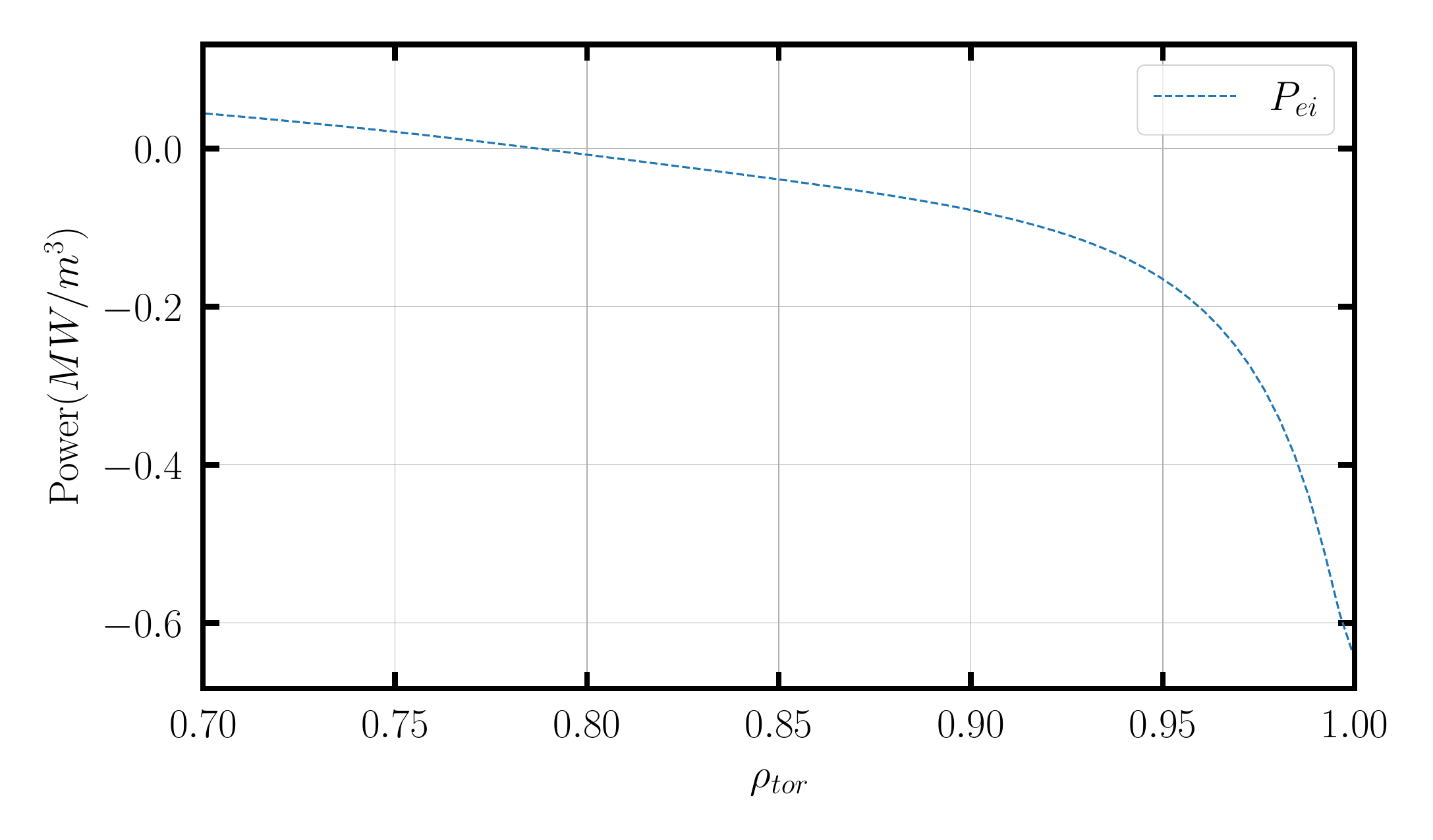}}{\rule{\linewidth}{6cm}}
        \caption{132588}
    \end{subfigure}
    \caption{$P_{ei}$ profiles for the evolving temperature case for (a) 132543 and (b) 132588. The energy exchange maintains the same dipole structure as the ETG-only case, with energy transferring to electrons in the steep gradient region ($P_{ei} < 0$) and to ions in the core ($P_{ei} > 0$).}
    \label{fig:pei_profile_neo}
\end{figure}

Returning to the fully coupled model defined in Equations \ref{eq:chie} and \ref{eq:chii}, the next configuration allows both species ($T_e$ and $T_i$) to evolve dynamically while keeping only the density fixed. 

Figure~\ref{fig:te_profile_neo} displays the temperature profile predictions under this coupled transport model utilizing the unscaled baseline ETG transport. While the profile shape remains intact, both temperature profiles are notably overpredicted relative to the experiment. The fact that $T_i$ and $T_e$ are overpredicted implies that the transport mechanisms in these simulations are not sufficiently comprehensive; the profiles are forced to steepen artificially to push out the required input power.  Note that the main difference between these simulations and the earlier simulations (electron channel only) is that the $T_i$ profile is also evolving and thus cannot act as an energy sink for the electrons, which effectively maintained closer proximity to the experimental profiles.  This illustrates the importance of \textit{integrated} modeling to develop a full understanding of pedestal transport and dynamics. 

As illustrated in Figure~\ref{fig:q_profile_neo}, the transport mechanisms exhibit distinct spatial profiles. Neoclassical transport provides a consistently large baseline across the entire region. In contrast, ETG turbulence is highly localized, providing substantial transport in the pedestal top and plasma edge where $\eta_e$ is large. Moving inward toward the core where the gradients flatten and $\eta_e$ drops, the ETG drive collapses, leaving the large baseline neoclassical transport as the primary thermal channel.

Crucially, it is evident that neoclassical transport itself cannot hold the ion temperature pedestal for 132543 and 132588. This shortfall is especially severe for 132588, where the modeled ion temperature vastly exceeds the experimental measurement due to insufficient thermal transport. Because $\chi_{i,\mathrm{neo}}$ forms the baseline transport for the ions in this configuration, its insufficiency points strongly to the conclusion that another anomalous ion transport channel is required to capture the full transport balance and pedestal shape. The $P_{ei}$ profiles (Figure~\ref{fig:pei_profile_neo}) maintain their familiar shape, confirming that the internal collisional energy transfer is not responsible for the elevated ion temperature profiles, but rather a universal lack of outward radial thermal transport.

\begin{figure}[ht]
    \centering
    \begin{subfigure}{0.48\textwidth}
        \IfFileExists{132543_T_nscan.pdf}{\includegraphics[width=\linewidth]{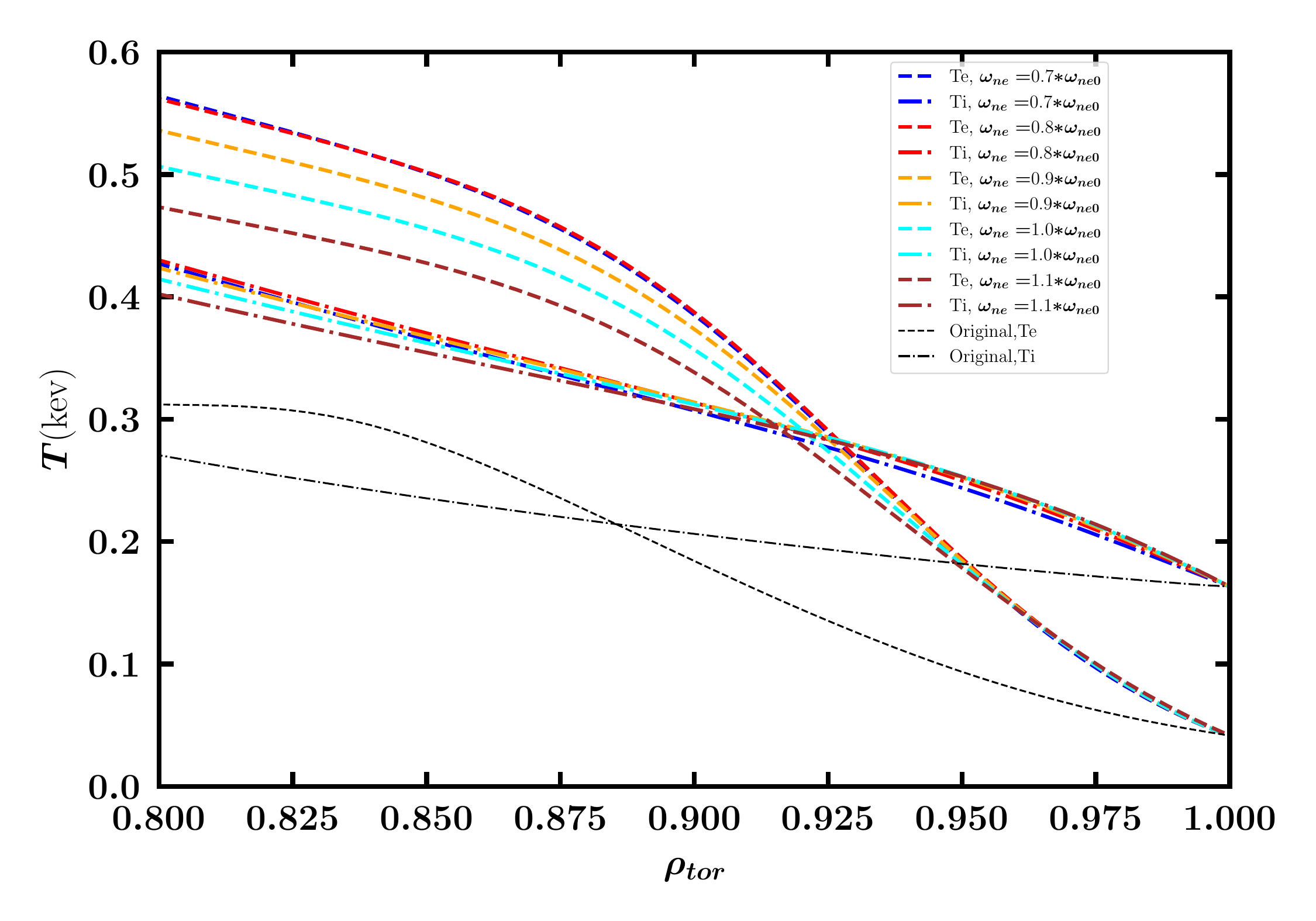}}{\rule{\linewidth}{6cm}}
        \caption{132543, pivot $\rho=1.0$}
    \end{subfigure}
    \hfill
    \begin{subfigure}{0.48\textwidth}
        \IfFileExists{132588_T_nscan.pdf}{\includegraphics[width=\linewidth]{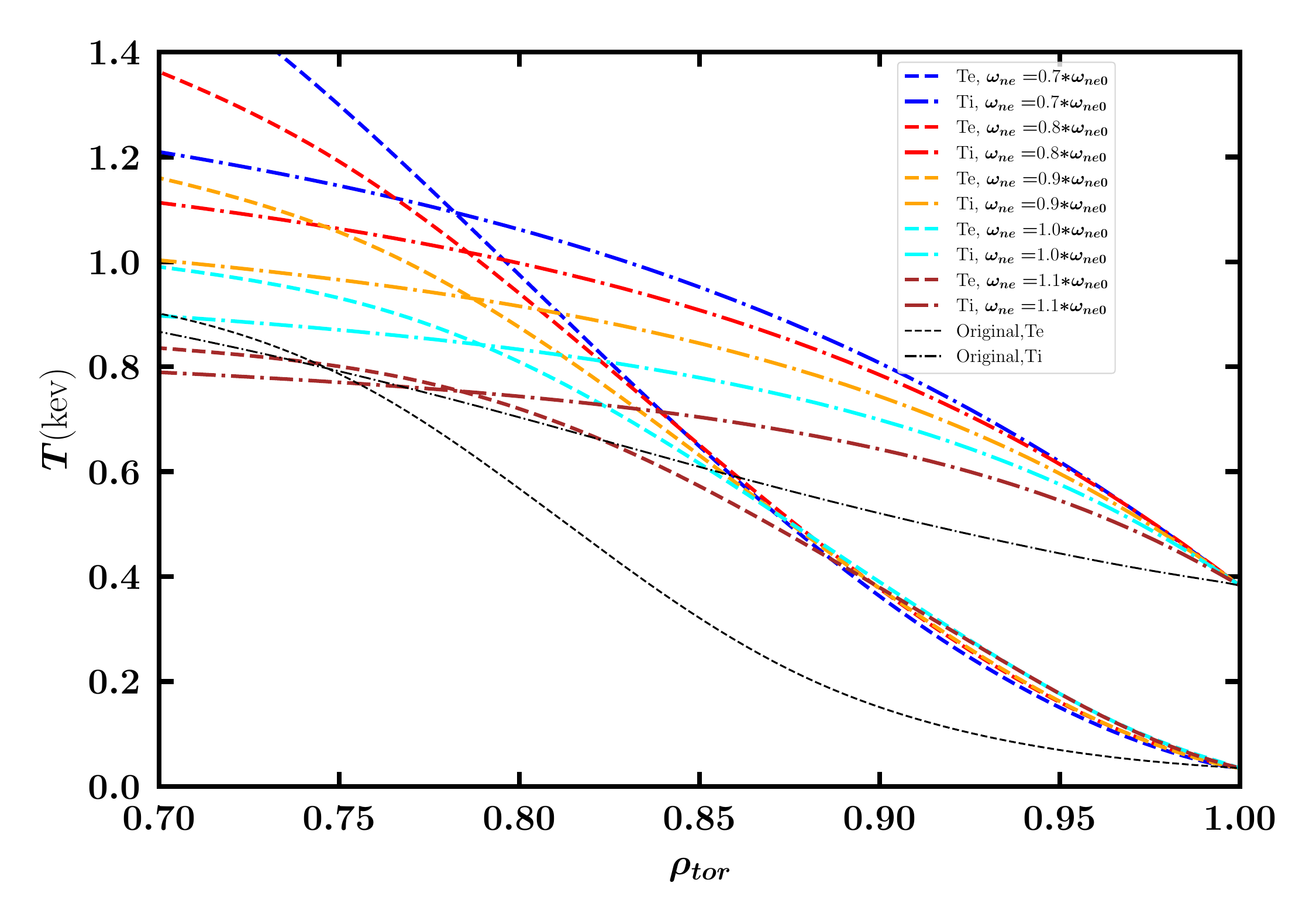}}{\rule{\linewidth}{6cm}}
        \caption{132588, pivot $\rho=1.0$}
    \end{subfigure}\\
    \vspace{0.5cm}
    \begin{subfigure}{0.48\textwidth}
        \IfFileExists{132543_T_nscan_nfix09.pdf}{\includegraphics[width=\linewidth]{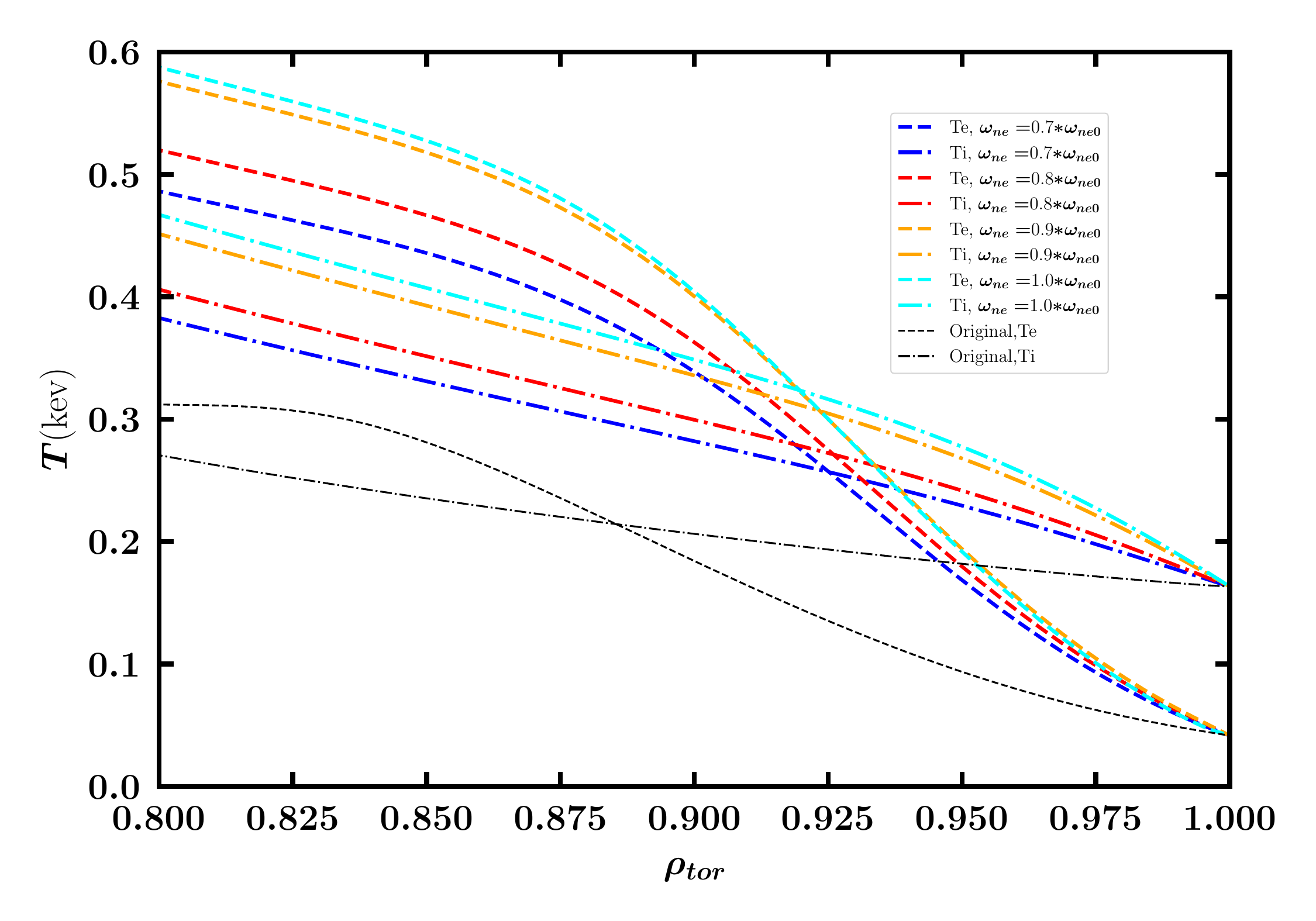}}{\rule{\linewidth}{6cm}}
        \caption{132543, pivot $\rho=0.9$}
    \end{subfigure}
    \hfill
    \begin{subfigure}{0.48\textwidth}
        \IfFileExists{132588_T_nscan_nfix09.pdf}{\includegraphics[width=\linewidth]{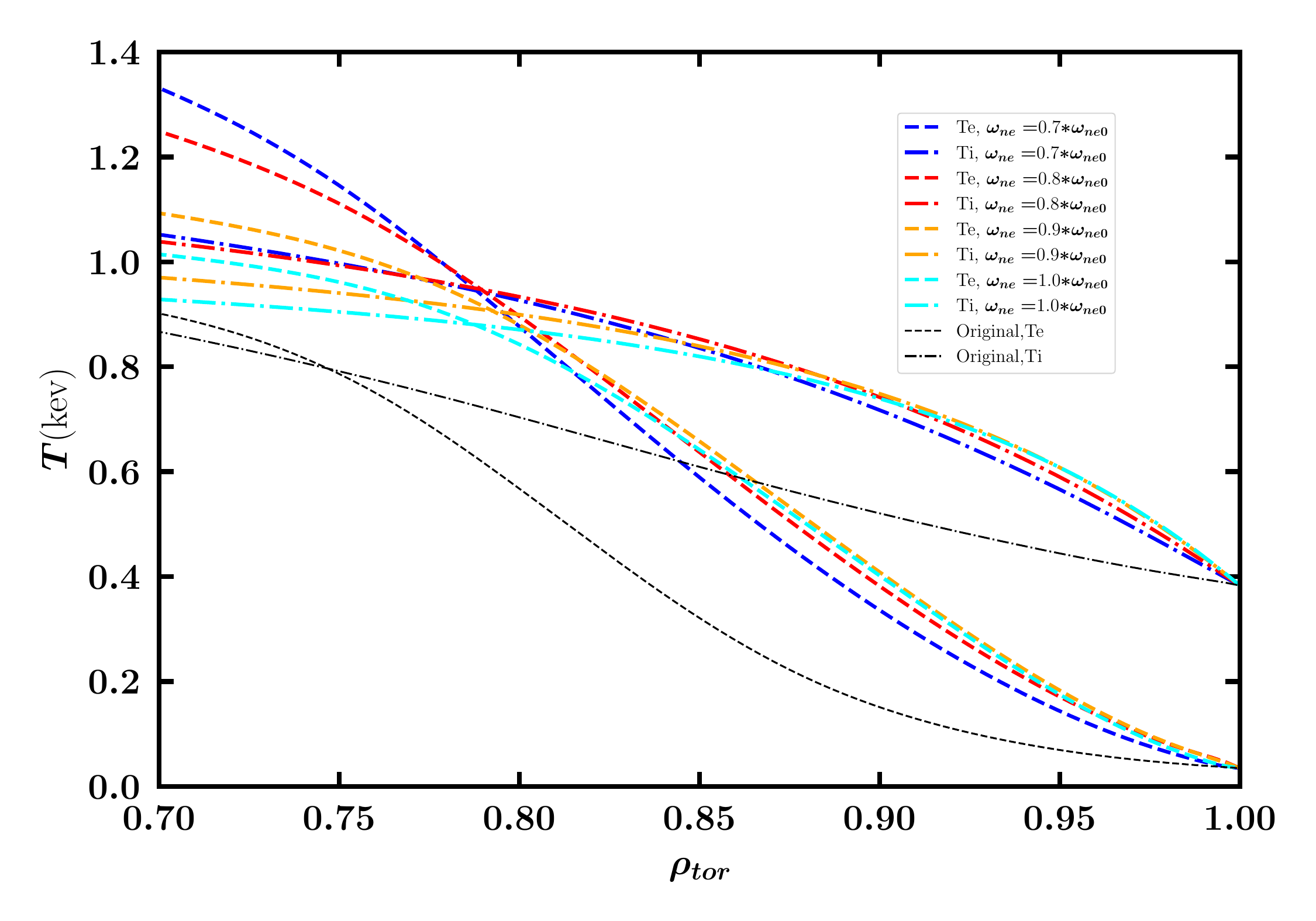}}{\rule{\linewidth}{6cm}}
        \caption{132588, pivot $\rho=0.9$}
    \end{subfigure}
    \caption{$T_e$ and $T_i$ profiles predicted with different density profiles (modifying density gradients by $-30\%$ to $+10\%$). Pivot points are at $\rho=1.0$ for (a) 132543 and (b) 132588, and at $\rho=0.9$ for (c) 132543 and (d) 132588. The dynamic evolution of $T_i$ does not break the model's resilience to density variations.}
    \label{fig:teti_dens_scan_neo}
\end{figure}

We again evaluate the sensitivity of the modeled profiles to adjustments in the density gradient, altering it by $-30\%$ to $+10\%$ at pivot points $\rho=1.0$ and $\rho=0.9$. As illustrated in Figure~\ref{fig:teti_dens_scan_neo}, allowing the ion temperature to evolve freely does not compromise the robustness of the system; neither the $T_e$ nor the $T_i$ profiles exhibit significant changes. As in the single-species case, the physical justification remains tied to the electron-ion energy exchange dynamics, where changes in the turbulent drive via $\eta_e$ are neutralized by corresponding density-dependent changes in the $P_{ei}$ collisional exchange.

\subsection{Impact of scaled ETG transport (\texorpdfstring{$2 \times Q_{\mathrm{ETG}}$}{2x QETG}) on profile evolution}

\begin{figure}[ht]
    \centering
    \begin{subfigure}{0.48\textwidth}
        \IfFileExists{132543_T2.pdf}{\includegraphics[width=\linewidth]{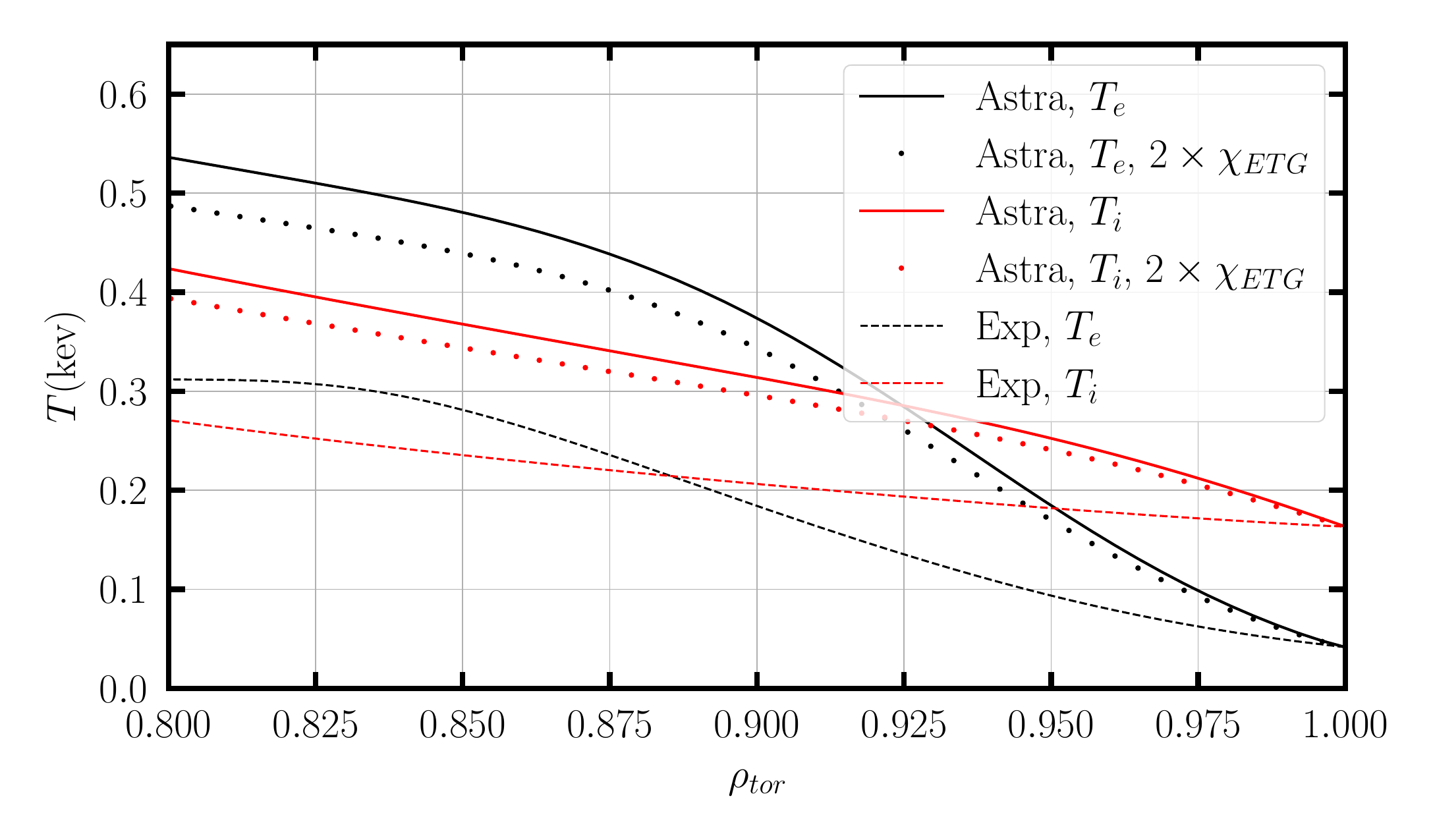}}{\rule{\linewidth}{6cm}}
        \caption{132543}
    \end{subfigure}
    \hfill
    \begin{subfigure}{0.48\textwidth}
      \IfFileExists{132588_T2.pdf}{\includegraphics[width=\linewidth]{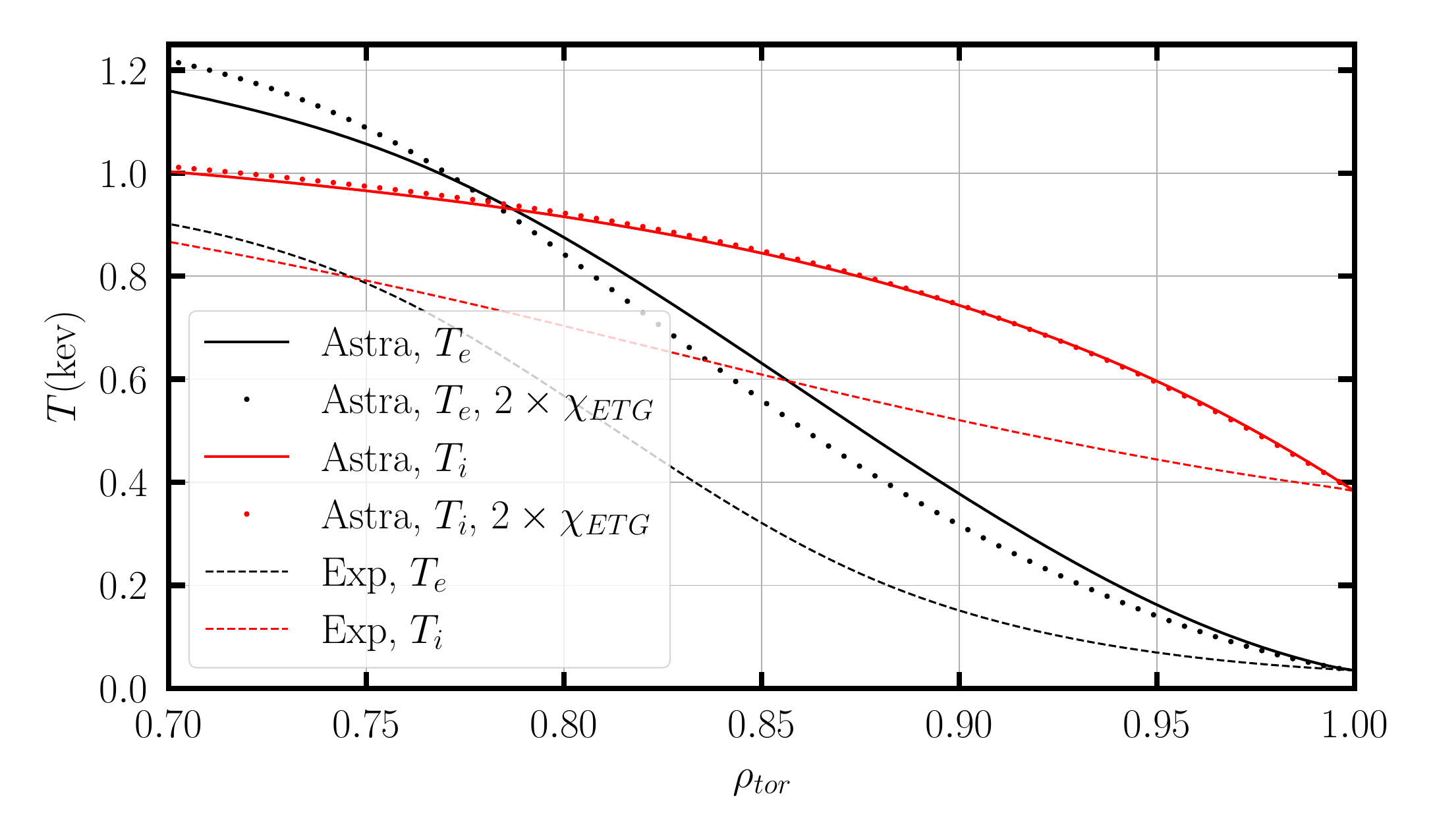}}{\rule{\linewidth}{6cm}}
        \caption{132588}
    \end{subfigure}
    \caption{Temperature profile predictions using the coupled two-species model, with the ETG transport manually scaled by a factor of two ($2 \times Q_{\mathrm{ETG}}$). While this scaling introduces slight, secondary adjustments to the profile shapes, the overall coupled simulation remains poor for both cases because the $T_i$ profiles are still heavily overpredicted by neoclassical limits.}
    \label{fig:two_spec_2x}
\end{figure}

In contrast to the single-species case where the effect from $P_{ei}$ strongly cancels the contribution from extra ETG transport, applying the $2\times$ scaling factor to the fully coupled two-species model ($\chi_e = 2 \times \chi_{e,\mathrm{ETG}} + \chi_{e,\mathrm{neo}}$) allows the temperatures to evolve somewhat differently. As demonstrated in Figure~\ref{fig:two_spec_2x}, the response to multiplying the ETG transport by two introduces minor adjustments to the profile shapes. However, for both discharges, the scaling only slightly modifies the macroscopic temperature profile and leaves the overall prediction severely elevated. 

Crucially, the modification fails to address the massive overprediction in the ion thermal channel. This divergence highlights a fundamental finding: a substantial improvement in the accuracy of the ETG transport with strong ion-electron thermal exchange cannot substitute for the clear physical lack of anomalous ion-scale transport at the pedestal top.

\section{Inclusion of KBM and MHD-like transport}
\label{sec:mhd_model}
To address the remaining shortfall in predicted heat flux and account for the missing ion transport channel identified by the failure of the ETG+Neoclassical model, we must consider additional instability mechanisms. It is known from previous studies that Kinetic Ballooning Modes (KBM) and other ion-scale MHD-like instabilities are frequently found in and can regulate the pedestal of spherical tokamaks like NSTX, including specifically discharges 132543 and 132588~\cite{Diallo2013, Guttenfelder2012, Kaye2013}.

\subsection{Quasi-linear surrogate model}
To systematically evaluate the stability boundaries of these modes within our experimental framework, we mapped out the stability space by artificially modifying the experimental $T_e$ and $T_i$ profiles in the range of $+10\%$ to $-20\%$. For each of these varied pressure profiles, a corresponding fully self-consistent MHD equilibrium was reconstructed using \textsc{astra}'s built-in equilibrium code, \textsc{spider}~\cite{Ivanov2005}, and subsequently refined using the \textsc{chease} code~\cite{Lutjens1996} to ensure consistent geometric coefficients. Linear gyrokinetic simulations were then performed using the \textsc{gene} code across this database of equilibria.

\begin{figure}[ht]
    \centering
    \begin{subfigure}{0.48\textwidth}
        \IfFileExists{kbm_thread_132543.pdf}{\includegraphics[width=\linewidth]{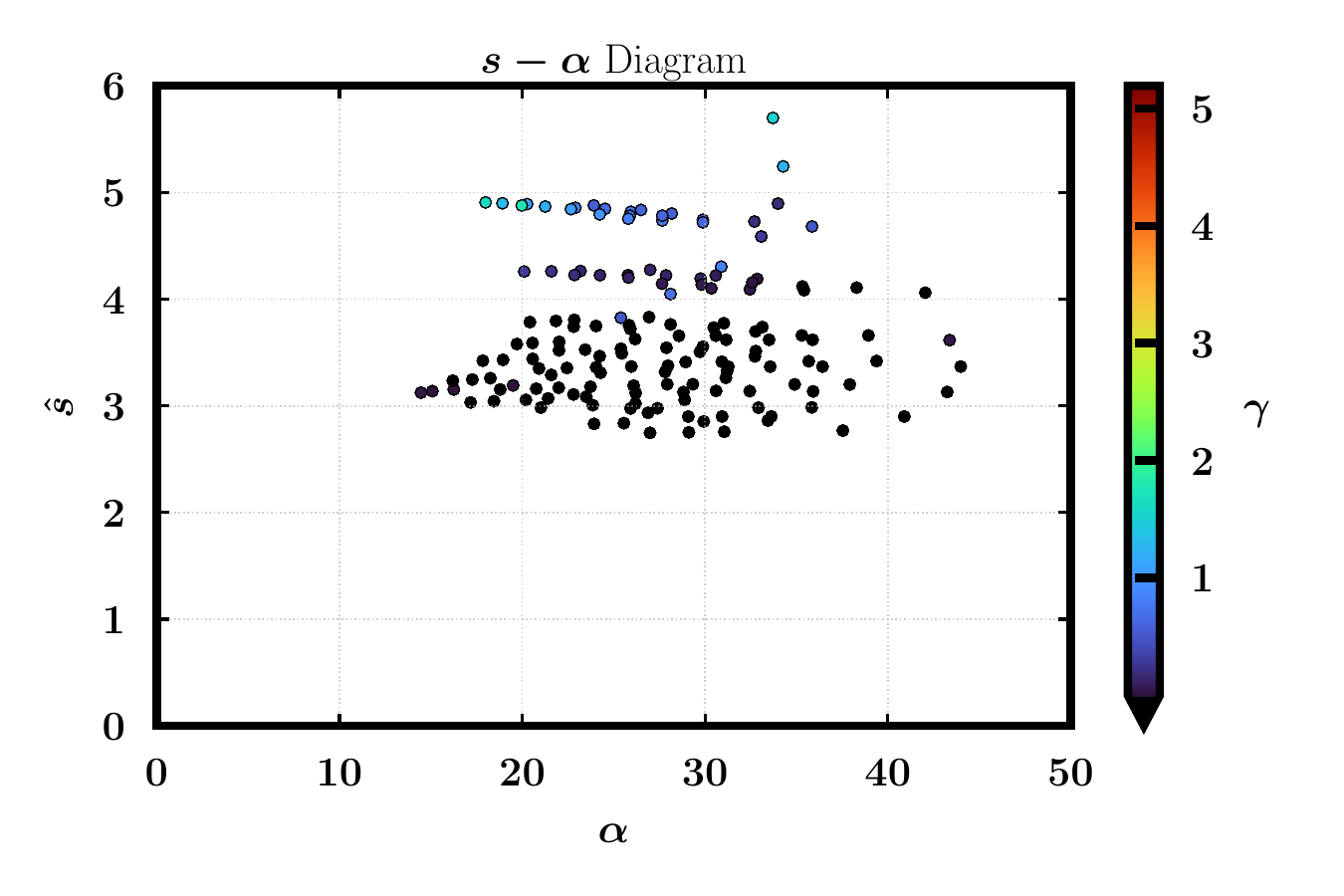}}{\rule{\linewidth}{6cm}}
        \caption{132543}
    \end{subfigure}
    \hfill
    \begin{subfigure}{0.48\textwidth}
        \IfFileExists{kbm_thread_132588.pdf}{\includegraphics[width=\linewidth]{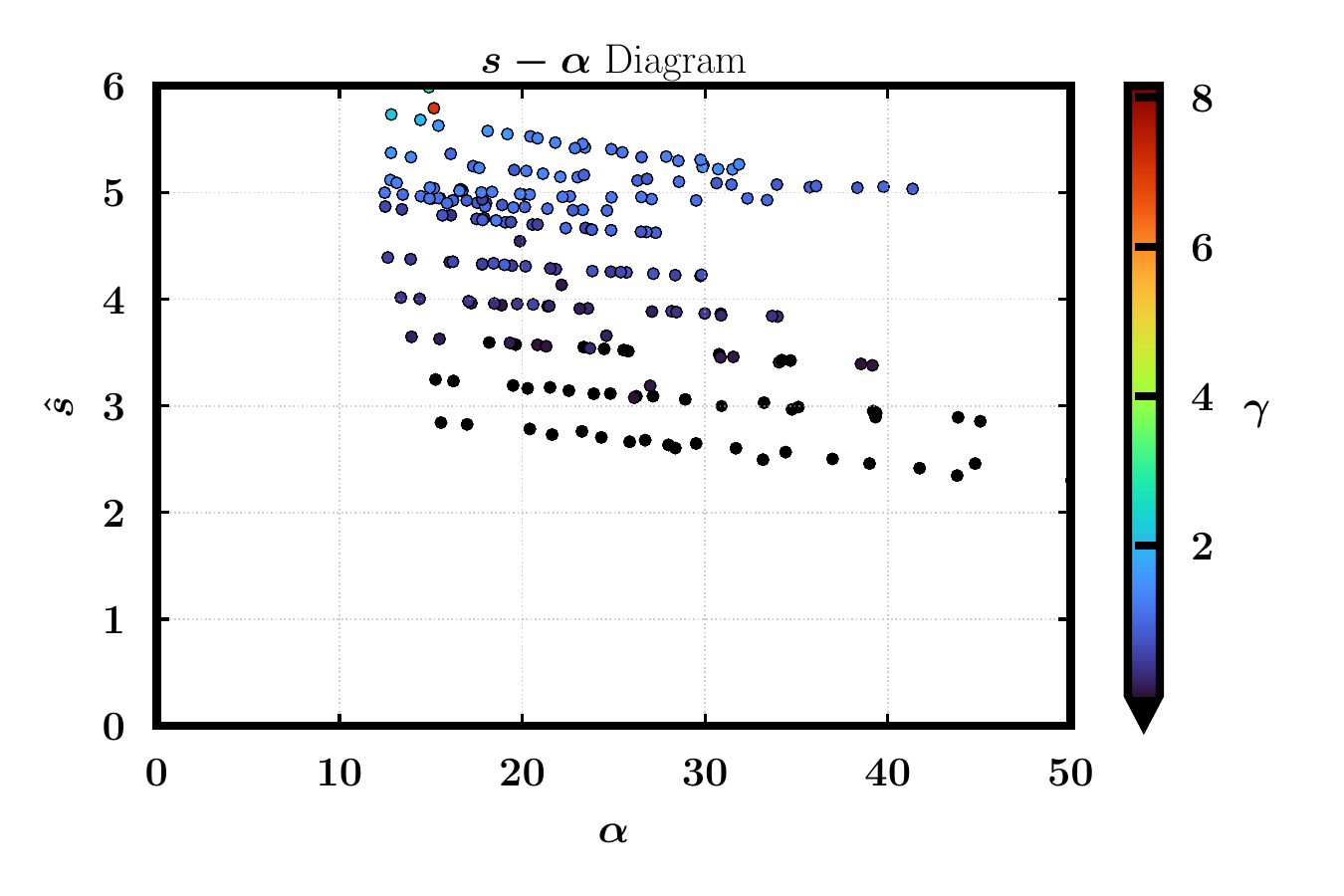}}{\rule{\linewidth}{6cm}}
        \caption{132588}
    \end{subfigure}
    \caption{Growth rate scattered heat plots on $s-\alpha$ diagrams at multiple radial locations for (a) 132543 and (b) 132588. The linear growth rates are calculated via \textsc{gene} gyrokinetic simulations based on augmented pressure profiles with self-consistent equilibria reconstructed using ASTRA (SPIDER) and CHEASE, scanning over global toroidal mode numbers $n_{0,\mathrm{global}} = 2, 4, 8, 16$. The results clearly illustrate the instability thresholds for KBM and MHD-like modes. (Simulations performed with the \textsc{gene} code).}
    \label{fig:s_alpha}
\end{figure}

Figure~\ref{fig:s_alpha} presents the resulting linear growth rates---obtained by scanning over the global toroidal mode numbers $n_{0,\mathrm{global}} \in \{2, 4, 8, 16\}$---displayed as scattered heat plots on $s-\alpha$ diagrams for both shots at multiple radial locations. Here, $\hat{s}$ is the magnetic shear and $\alpha = - \frac{2 \mu_0 R q^2}{B^2} \frac{dp}{dr}$ is the normalized pressure gradient. The parameter scans across the $s-\alpha$ diagrams were performed by varying the experimental temperature gradients within their standard error bars. This effectively maps out the KBM and MHD-like mode stability thresholds, clearly illustrating that the experimental pedestals operate directly within or around these critical limits.

Given that the experimental profiles sit near these stability boundaries, the KBM and MHD-like modes likely provides the physical mechanism responsible for constraining the pedestal profiles. To dynamically account for this transport in our predictive modeling, we introduce a quasi-linear surrogate model. Following the approach utilized by Giacomin \textit{et al.}~\cite{Giacomin2025}, the effective thermal diffusivity associated with these modes is expressed as:
\begin{equation}
\chi_{\mathrm{mix}} = c_0 \max_{k_y} \frac{\gamma}{\langle k_\perp^2 \rangle_{\chi}},
\label{eq:MHDmodel}
\end{equation}
where $c_0$ is an empirical scaling coefficient, $k_y$ is the binormal wavenumber, and $\gamma$ is the mode growth rate derived from the stability analysis. Unlike standard eigenmode averaging, the perpendicular wavenumber $k_\perp$ here is evaluated as a weighted sum over all fields ($\chi = \phi, A_\parallel, B_\parallel$) to serve as a generalized field variable, a form shown to better capture the transport scaling~\cite{Giacomin2025}. Furthermore, our evaluation of the $k_\perp$ average incorporates an additional Bessel function weighting to accurately capture finite Larmor radius (FLR) effects, which effectively reduces the impact of long tails in the eigenmodes.  

To efficiently evaluate this diffusivity within the \textsc{astra} transport solver, an automated Python routine is utilized. This routine is designed to identify the type of dominant micro-instability present and then calculate the appropriate quasi-linear model for that specific mode. For this study, the routine was used to construct a surrogate model via a Radial Basis Function (RBF) interpolation. Configured with a linear kernel and a polynomial degree of 1 to ensure robust linear extrapolation, this RBF interpolator is trained on the comprehensive database of \textsc{gene} gyrokinetic simulations generated for the $s-\alpha$ stability scans shown in Figure~\ref{fig:s_alpha}. While ETG turbulence is generally restricted to regions of small density gradients ($\eta_e \gg 1$), ion-scale KBM and MHD-like modes are driven by the total pressure gradient ($\alpha$). This allows KBMs to thrive in large density gradient regions, providing a highly complementary transport channel.

The surrogate model evaluates the quasi-linear transport as a continuous function of the local ion temperature ($T_i$), electron temperature ($T_e$), and magnetic shear ($\hat{s}$). To interface cleanly with the transport solver, the Python RBF routine generates a structured JSON file containing the modeled transport coefficients. \textsc{astra} then reads this JSON file to continuously extract the appropriate numbers for the thermal diffusivities (specifically the $\chi_i$ and $\chi_e$ contributions from the KBM/MHD-like modes) at each time step during its profile evolution. The magnitude of this transport is calibrated globally by setting the empirical prefactor to $c_0 = 0.0008$ based on benchmark comparisons.  This effectively supplies a surrogate model for KBM transport for a space of equilibria in close proximity to the experimental NSTX equilibrium of interest.  Ongoing work is establishing a broader database and corresponding surrogate model for pedestal transport, which can be seamlessly incorporated in these workflows in the future. 

\begin{figure}[ht]
    \centering
    \begin{subfigure}{0.48\textwidth}
        \IfFileExists{132543_KBM_T+-10.pdf}{\includegraphics[width=\linewidth]{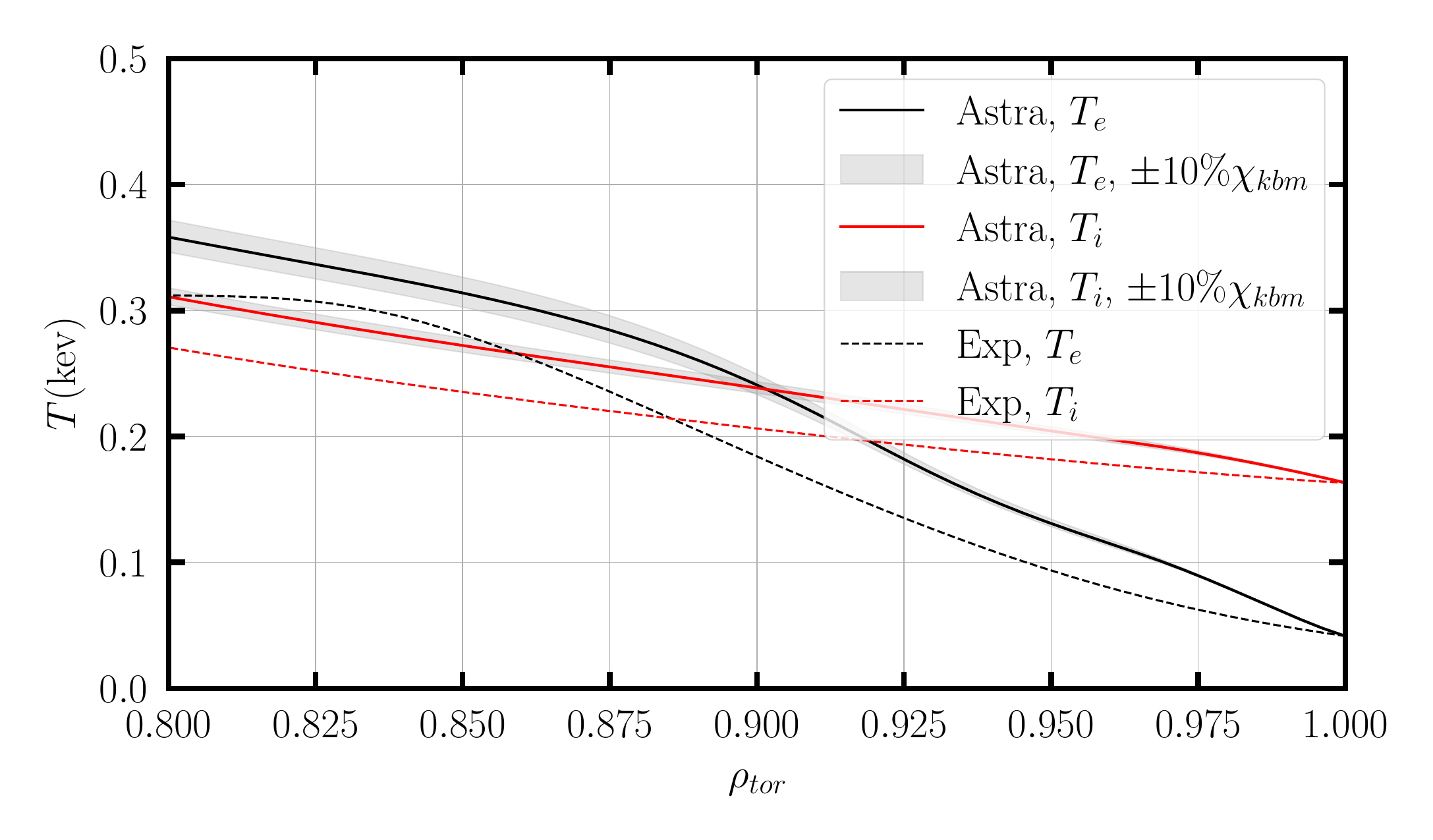}}{\rule{\linewidth}{6cm}}
        \caption{132543}
    \end{subfigure}
    \hfill
    \begin{subfigure}{0.48\textwidth}
        \IfFileExists{132588_KBM_T+-10.pdf}{\includegraphics[width=\linewidth]{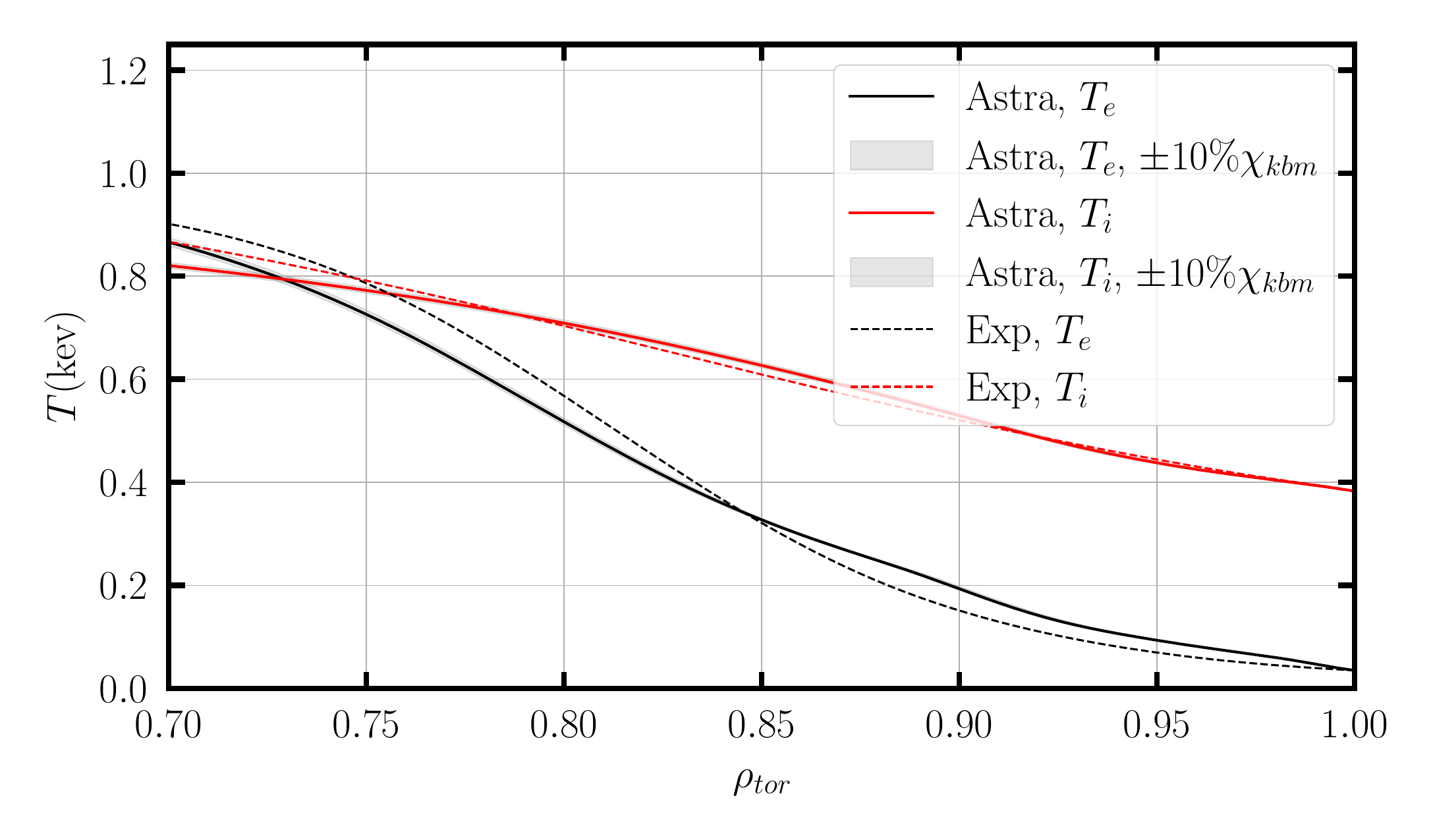}}{\rule{\linewidth}{6cm}}
        \caption{132588}
    \end{subfigure}
    \caption{Temperature profile predictions incorporating the quasi-linear surrogate model for KBM/MHD-like transport alongside the unscaled baseline ETG and neoclassical channels. The shaded bands indicate a $\pm 10\%$ variation applied to $\chi_{\mathrm{KBM}}$, which results in little overall variance, primarily affecting the pedestal top rather than the steep gradient region or plasma edge.}
    \label{fig:surrogate_profiles}
\end{figure}

\begin{figure}[ht]
    \centering
    \begin{subfigure}{0.48\textwidth}
        \IfFileExists{132543_KBM_chi.pdf}{\includegraphics[width=\linewidth]{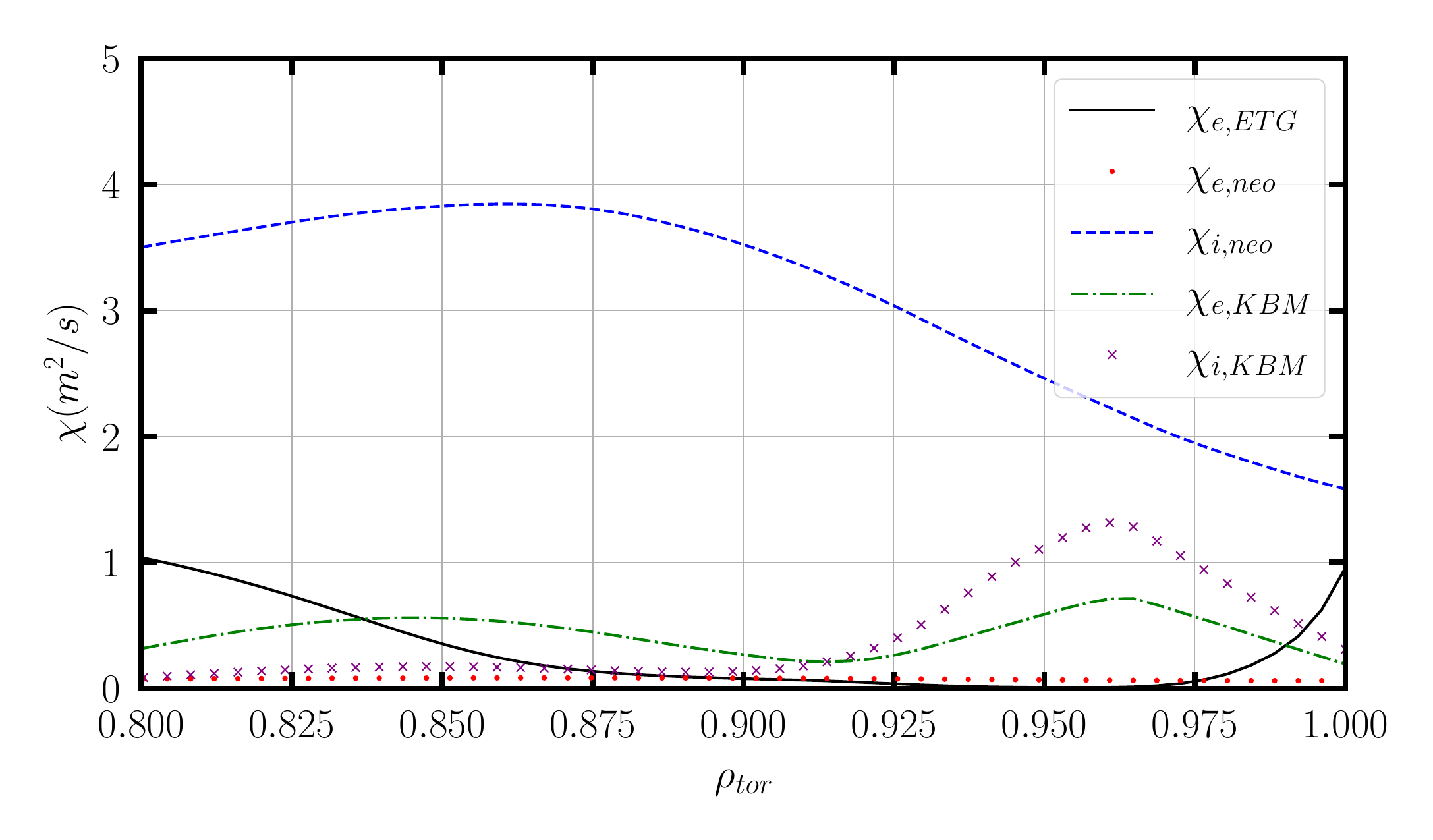}}{\rule{\linewidth}{6cm}}
        \caption{132543}
    \end{subfigure}
    \hfill
    \begin{subfigure}{0.48\textwidth}
        \IfFileExists{132588_KBM_chi.pdf}{\includegraphics[width=\linewidth]{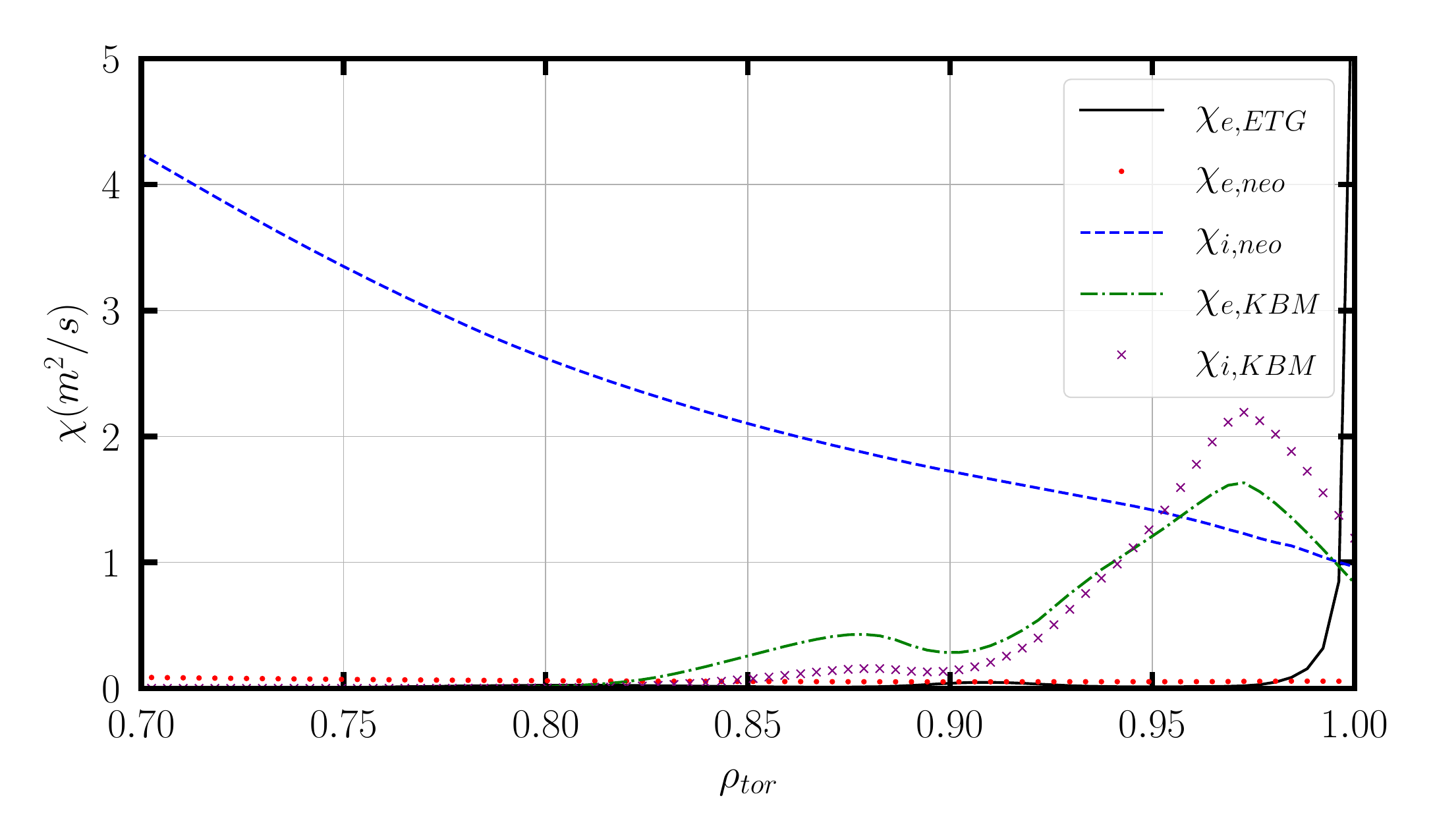}}{\rule{\linewidth}{6cm}}
        \caption{132588}
    \end{subfigure}
    \caption{Radial profiles of the thermal diffusivities for the coupled simulations incorporating the KBM/MHD-like surrogate model for (a) 132543 and (b) 132588. The plots illustrate the individual contributions to the total transport: $\chi_{i,\mathrm{KBM}}$, $\chi_{i,\mathrm{neo}}$, $\chi_{e,\mathrm{ETG}}$, and $\chi_{e,\mathrm{KBM}}$. (Simulations performed with the \textsc{gene} code).}
    \label{fig:kbm_chi}
\end{figure}

The total thermal diffusivities for each species then become:
\begin{align}
\chi_e &= \chi_{e,\mathrm{ETG}} + \chi_{e,\mathrm{neo}} + f_e \chi_{\mathrm{mix}},\\
\chi_i &= \chi_{i,\mathrm{neo}} + f_i \chi_{\mathrm{mix}},
\end{align}
where $f_e$ and $f_i$ are partition factors describing how the MHD-like mode couples to the electron and ion channels, respectively. Incorporating this extracted data within \textsc{astra} alongside the unscaled ETG baseline yields improved agreement with experiment (Figure~\ref{fig:surrogate_profiles}). To assess the sensitivity of these results, the KBM transport diffusivity ($\chi_{\mathrm{KBM}}$) was varied by $\pm 10\%$. As depicted by the narrow shaded regions in Figure~\ref{fig:surrogate_profiles}, this results in little overall variance in the final temperature profiles. Notably, this variation alters the temperature profile more at the pedestal top, while having only a small effect on the steep gradient region and the plasma edge. This highlights that the ion-scale transport is stiff; the profiles are tightly constrained by the marginal stability boundary of the KBM/MHD-like modes. To further illustrate the underlying transport balance, Figure~\ref{fig:kbm_chi} visualizes the corresponding thermal diffusivities, delineating the distinct contributions from $\chi_{i,\mathrm{KBM}}$, $\chi_{i,\mathrm{neo}}$, $\chi_{e,\mathrm{ETG}}$, and $\chi_{e,\mathrm{KBM}}$ across the pedestal. 

As shown in Figure~\ref{fig:surrogate_profiles}(b), the inclusion of the quasi-linear surrogate model produces a good match to the experimental data for discharge 132588. This agreement is expected, as 132588 is an ELM-free discharge where the pedestal reaches a true steady state, making the steady-state solver results unambiguous. However, for discharge 132543 (Figure~\ref{fig:surrogate_profiles}(a)), the modeled temperatures remain slightly overpredicted. This discrepancy is physically consistent with the experimental scenario: as noted in Section~\ref{sec:ETG_only}, the experimental data and the corresponding profile used for analysis for 132543 are at time 700~ms, during a post-ELM recovery phase. Because the experimental temperature profiles are still slowly growing and have not yet fully saturated, the steady-state \textsc{astra} predictions inherently overshoot the transient experimental measurements.

\subsection{Model with \texorpdfstring{$E \times B$}{E x B} shear suppression}
In the steep gradient region of the pedestal, strong radial electric fields generate substantial $E \times B$ flow shear, which is known to suppress turbulent transport. To further refine the quasi-linear surrogate model, an explicit $E \times B$ shear suppression factor, $\Theta_{E \times B}$, was introduced. The radial electric field ($E_r$) utilized in these evaluations is evaluated with the neoclassical radial force balance equation.
    
Following the first-principles theoretical framework developed by Hatch \textit{et al.}~\cite{Hatch2018}, this suppression factor is formulated as:
\begin{equation}
\Theta_{E \times B} = \left( 1.0 + \frac{k_y}{\langle k_x \rangle} \frac{\gamma_{E \times B}}{\gamma} \right)^{-2},
\end{equation}
where $\gamma_{E \times B}$ is the $E \times B$ shearing rate and $\gamma$ is the linear mode growth rate. The average radial wavenumber, $\langle k_x \rangle$, is approximated by:
\begin{equation}
\langle k_x \rangle \approx \hat{s} k_y w,
\end{equation}
where $w$ represents the average mode width. This width is calculated as a weighted average of the mode structures across the relevant fluctuating fields (e.g., the electrostatic potential $\phi$, and the parallel vector potential components $A_\parallel$ and $B_\parallel$, depending on the simulation parameters), normalized by the maximum field amplitudes. 

\begin{figure}[ht]
    \centering
    \begin{subfigure}{0.48\textwidth}
        \IfFileExists{132543_KBM_Texb.pdf}{\includegraphics[width=\linewidth]{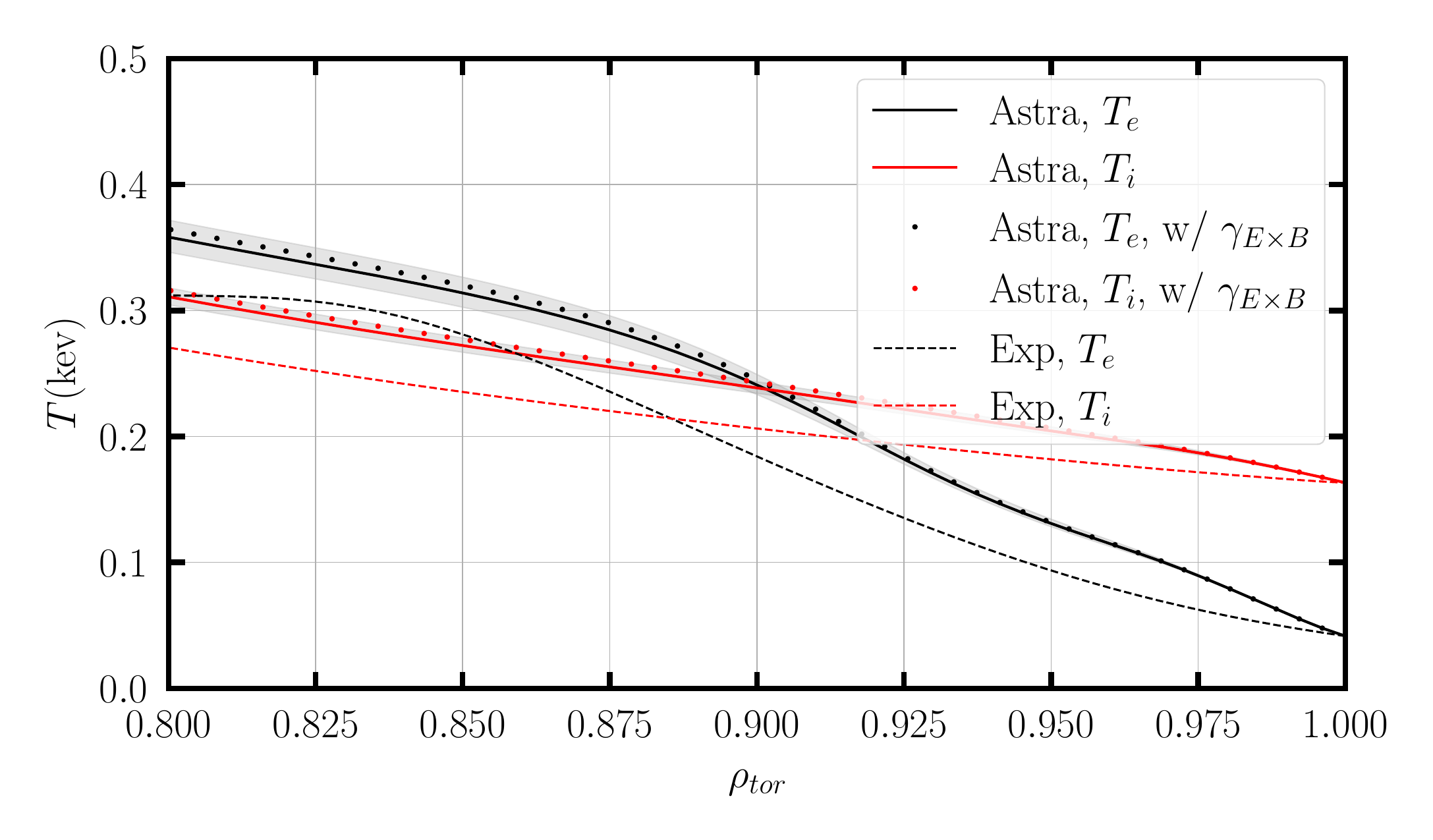}}{\rule{\linewidth}{6cm}}
        \caption{132543}
    \end{subfigure}
    \hfill
    \begin{subfigure}{0.48\textwidth}
       \IfFileExists{132588_KBM_Texb.pdf}{\includegraphics[width=\linewidth]{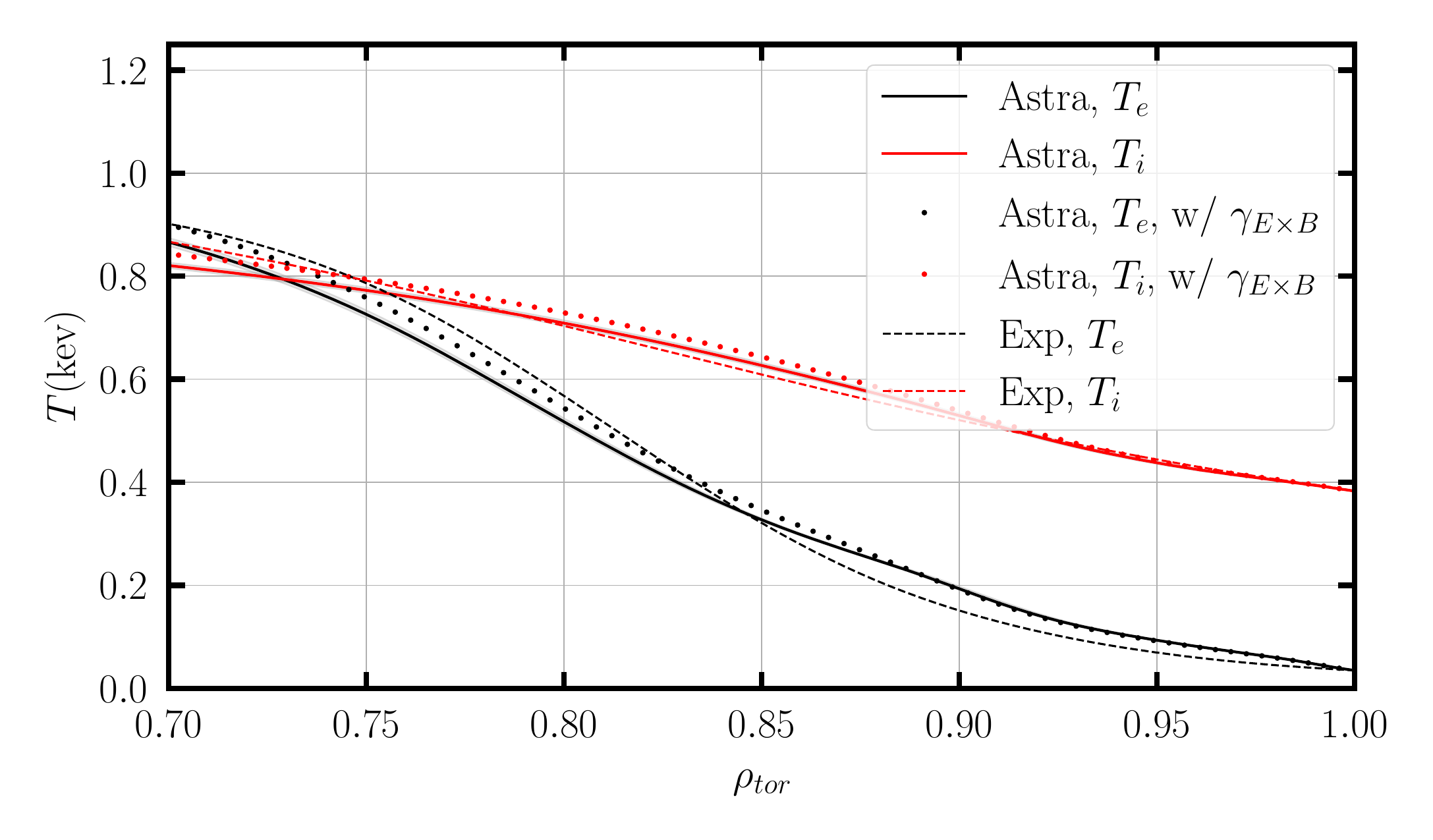}}{\rule{\linewidth}{6cm}}
        \caption{132588}
    \end{subfigure}
    \caption{Temperature profile predictions using the surrogate transport model augmented with the $E \times B$ shear suppression factor $\Theta_{E \times B}$. While the absolute temperature magnitudes are slightly elevated compared to the unsuppressed case, the explicit shear suppression introduces minor refinements that alter the overall curvature of the temperature profiles.}
    \label{fig:exb_suppression}
\end{figure}

In these evaluations, the calculated suppression factor $\Theta_{E \times B}$ is typically around 0.9 in the pedestal region, dropping as low as roughly 0.8 in the steepest gradient zones. Applying this factor directly to the mixed diffusivity ($\chi_{\mathrm{mix}} \rightarrow \Theta_{E \times B} \chi_{\mathrm{mix}}$) regulates the predicted transport in the presence of strong sheared flows. However, as demonstrated in Figure~\ref{fig:exb_suppression}, the inclusion of $E \times B$ shear suppression does not lower the absolute magnitudes of the modeled $T_e$ and $T_i$ profiles; in fact, the predicted temperatures for 132543 are slightly higher and marginally worse in terms of absolute magnitude compared to the unsuppressed case. Because the transport driven by KBM and MHD-like modes is stiff, a 10\% to 20\% reduction in the quasi-linear transport coefficient is easily compensated for by an almost imperceptible steepening of the pressure gradients. Instead, its primary effect is slightly modifying the temperature profiles, helping to introduce a small but noticeable refinement to the overall shape and curvature of the modeled data.

\subsection{Full transport model with scaled ETG (\texorpdfstring{$2 \times Q_{\mathrm{ETG}}$}{2x QETG}) and MHD surrogate}
To explicitly test the combined impact of our findings and construct a robust representation of the NSTX pedestal, we finalize our modeling by combining the $2 \times \chi_{e,\mathrm{ETG}}$ scaled model with the coupled KBM/MHD-like surrogate model and standard neoclassical transport. The empirical prefactor for the KBM/MHD surrogate is fixed at $c_0 = 0.0008$ (calibrated to discharge 132588) across both cases.

\begin{figure}[ht]
    \centering
    \begin{subfigure}{0.48\textwidth}
        \IfFileExists{132543_KBM_T2.pdf}{\includegraphics[width=\linewidth]{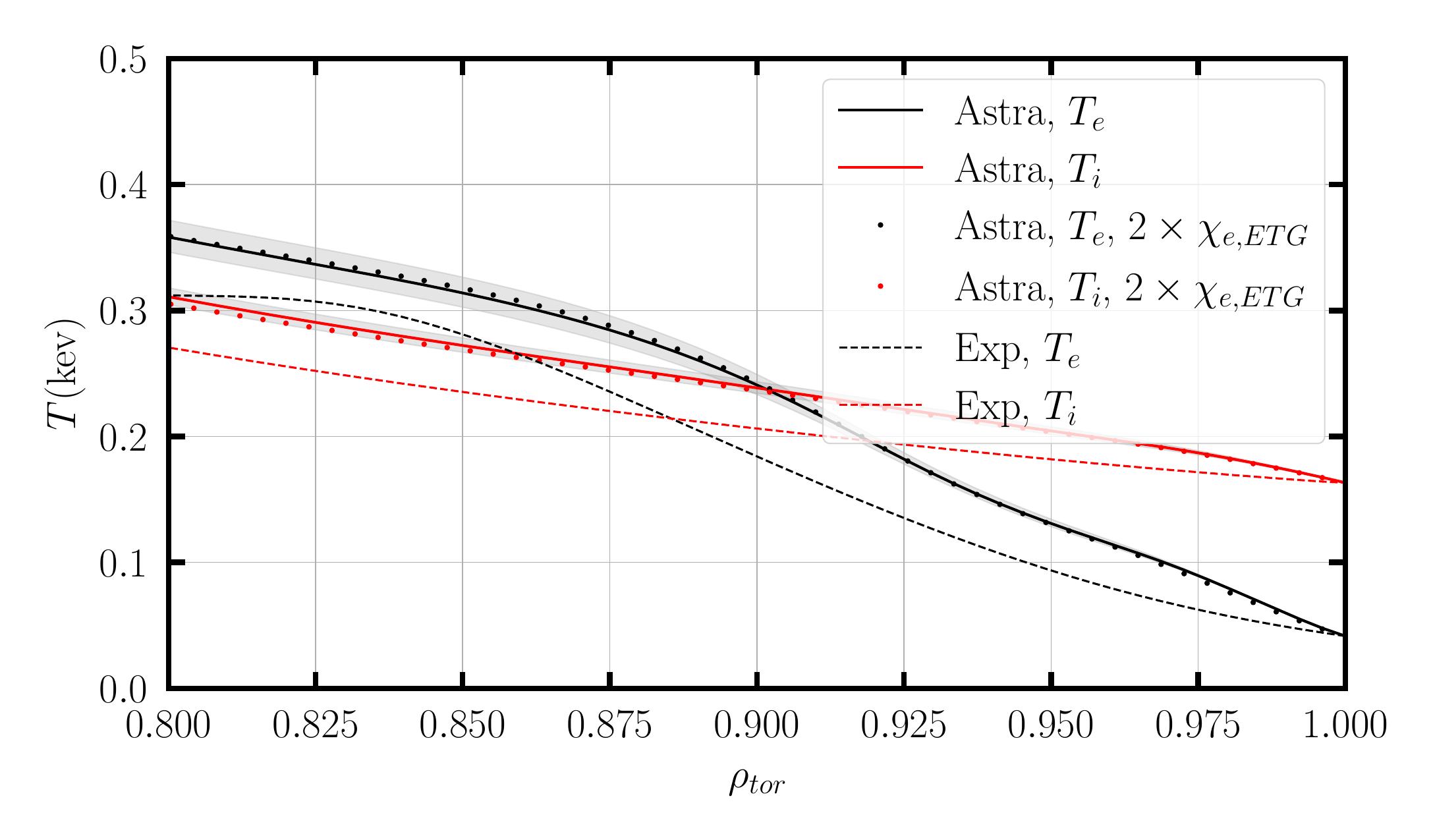}}{\rule{\linewidth}{6cm}}
        \caption{132543}
    \end{subfigure}
    \hfill
    \begin{subfigure}{0.48\textwidth}
        \IfFileExists{132588_KBM_T2.pdf}{\includegraphics[width=\linewidth]{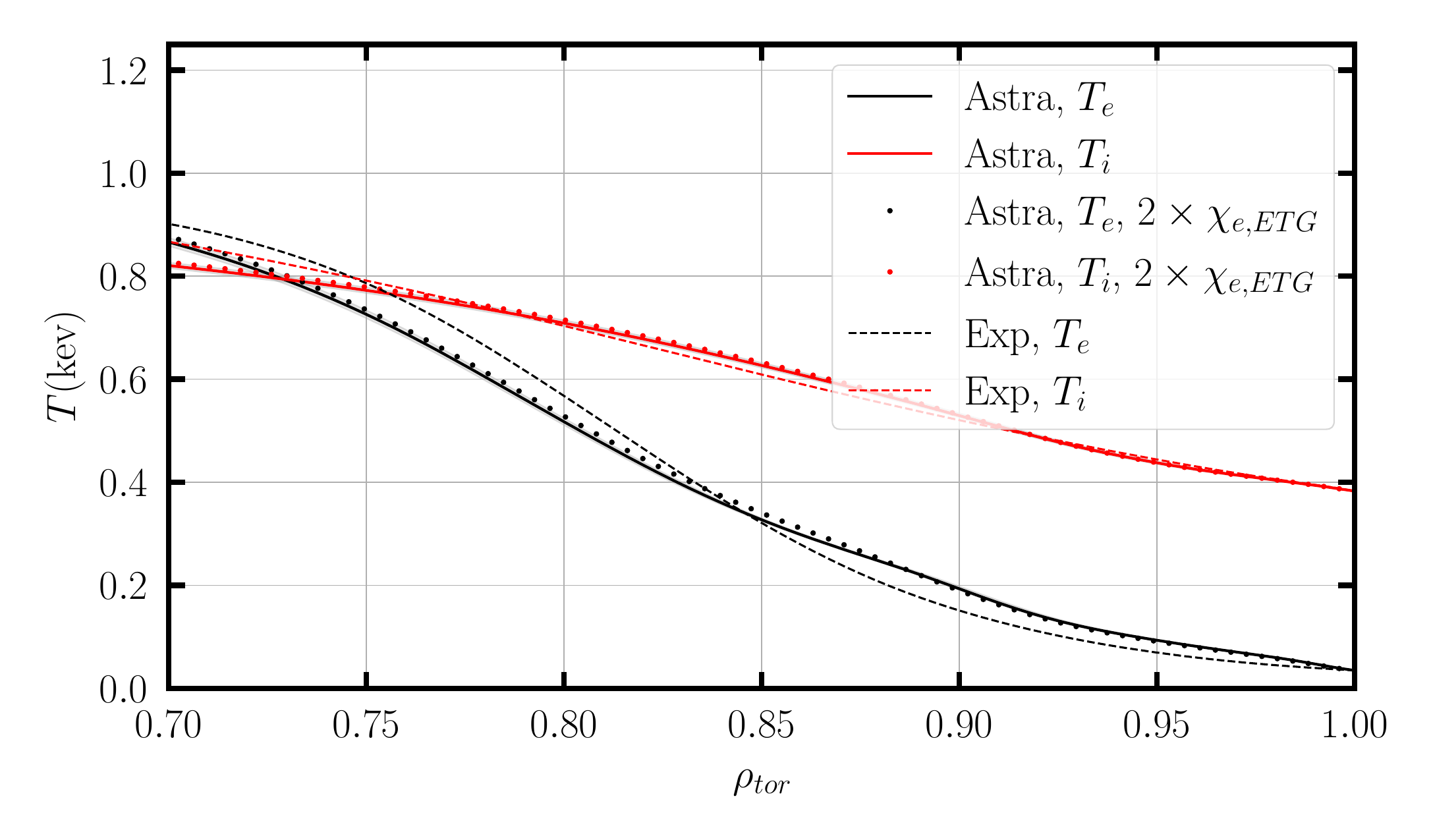}}{\rule{\linewidth}{6cm}}
        \caption{132588}
    \end{subfigure}
    \caption{Final temperature profile predictions combining the KBM/MHD-like surrogate model (using a fixed prefactor of $c_0=0.0008$) with the $2 \times Q_{\mathrm{ETG}}$ scaled electron transport. This combination produces a strong match to the experimental data for both discharges.}
    \label{fig:full_transport_2x}
\end{figure}

As shown in Figure~\ref{fig:full_transport_2x}, this integrated model yields a good match to the experimental data for both discharges 132543 and 132588. The combination correctly balances the power flows across both channels, successfully relaxing both the $T_e$ and $T_i$ profiles to near experimental levels simultaneously. This combination highlights the spatial and parametric complementarity of the modes. Provided the temperature gradient is sufficiently large to drive transport, ETG transport dominates where the local density gradient is small (resulting in $\eta_e \gg 1$), while ion-scale KBM/MHD-like modes regulate regions with large total pressure gradients, allowing them to provide transport in areas with steep density gradients. Thus, the transport balance across the pedestal is fundamentally determined by the density profile shape. While the ion-scale MHD mode provides the critical anomalous bulk transport necessary to bring the overall temperature magnitudes down to experimental levels, it is important to note that neoclassical transport also remains a consistently large mechanism in impeding the growth of the pedestal, especially for the ion temperature profile. Meanwhile, the $2\times$ ETG scaling acts as a subtle modifier. Similar to the $E \times B$ shear suppression, though to an even lesser extent, it does not drastically change the absolute temperature magnitudes but provides a slight adjustment to the detailed shape of the profiles. This confirms that the accuracy of the electron thermal channel is highly sensitive to providing the correct bulk anomalous transport in the ion channel, and that the minor transport corrections are fully complementary.

\section{Conclusion}
\label{sec:conclusion}
In this study, a combination of reduced modeling approaches was implemented to predict pedestal profiles and investigate the underlying transport mechanisms governing the H-mode pedestal temperature profiles in NSTX discharges 132543 and 132588. Using the comprehensive 1.5D \textsc{astra} transport solver, we systematically assessed the roles of ETG turbulence, neoclassical physics, and ion-scale MHD-like instabilities in determining the pedestal profiles while keeping the pedestal density fixed.

A simple formula for ETG transport~\cite{Hatch2024} was compared with nonlinear simulations and found to underpredict them by a factor of two for these NSTX scenarios. Modifying the ETG transport model to compensate for this resulted in quantitative but not qualitative changes to the $T_e$ profile predictions. Initial investigations modeling the pedestal with ETG alone (using fixed ion temperature and density) successfully reproduced the qualitative shape of the electron temperature pedestal. However, coupling both species dynamically with only ETG and neoclassical transport revealed a severe deficit in anomalous ion transport, leading to massively overpredicted ion temperatures. To address this, a quasi-linear surrogate model for ion-scale instabilities (such as KBM) was developed based on linear \textsc{gene} simulations and coupled to \textsc{astra}.

One free parameter ($c_0 = 0.0008$) was calibrated to match the experimental profiles of discharge 132588. All other aspects of the model were directly determined by a quasilinear treatment of the underlying gyrokinetic simulations. The resulting transport was mapped throughout the relevant parameter space using \textsc{gene} simulations performed on a sequence of equilibria, which were self-consistently reconstructed using \textsc{astra} (SPIDER) and CHEASE on an ensemble of temperature and density profiles. Utilizing an automated routine to identify the instabilities and calculate the corresponding quasilinear transport, we incorporated a surrogate model that was coupled to \textsc{astra} to evolve pedestal profiles.

Ultimately, combining this ion-scale surrogate model along with the ETG model and neoclassical transport, \textsc{astra} simulations produced good agreement across both thermal channels for both discharges. Sensitivity scans varying the KBM transport magnitude by $\pm 10\%$ resulted in little overall variance in the predicted profiles, primarily altering the temperature at the pedestal top while having only a small effect on the steep gradient region and plasma edge, demonstrating that the turbulence is stiff. The modified twofold ETG transport provided only small secondary refinements to the final profile shape. The effect of $E \times B$ shear suppression was implemented in the quasilinear model and also resulted in only minor quantitative changes. One key insight from this work is that transport analysis of fixed profiles has substantial limitations since thermal exchange between the species ($P_{ei}$) can be a large effect, meaning coupled modeling of both channels is required.

In conclusion, our results indicate that balancing these mechanisms is essential to fully regulate the pedestal. Neoclassical transport is huge across the whole pedestal region for the ion transport channel. Provided the temperature gradient is sufficiently large to drive transport, the turbulent spatial domains are fundamentally determined by the local pedestal density profile. ETG is large in the plasma edge and low density gradient region ($\eta_e \gg 1$), which contributes substantially to the electron channel. Conversely, ion-scale KBM and MHD-like modes are also large and contribute to both ion and electron channels, driven by the large total pressure gradient in steep density regions. Ultimately, balancing the power flow across all these mechanisms is required to determine the final pedestal temperature profiles in spherical tokamaks like NSTX. Future efforts will focus on replacing the heuristic surrogate components with direct, physics-based quasilinear closures for KBM transport and evaluating MTM-driven transport. However, it is anticipated that MTM is a secondary effect after ETG and KBM, since these two primary mechanisms have already been shown to capture most of the transport effects. These continued developments will enable robust predictive modeling for NSTX-U and other future burning plasma experiments.

An important aspect of the present work is that, although the model is reduced, it remains strongly connected to first-principles physics. The framework contains only a single free parameter, which was calibrated using one experimental discharge yet successfully reproduced two discharges with very different pedestal characteristics. This suggests that the approach captures the dominant transport physics governing the NSTX pedestal.  

Several directions for future work follow naturally from this study. Further validation across a broader region of parameter space will be important, including comparisons with other spherical tokamaks such as MAST. Additional isolated gyrokinetic analyses will also help clarify the detailed roles of the various instabilities contributing to pedestal transport. The framework developed here is particularly well suited for predictive studies of future devices, including NSTX-U and STEP. Of particular interest are low-recycling scenarios in NSTX-U, where the present modeling capability may help capture the impact of separatrix conditions on pedestal confinement, potentially providing a missing link between edge boundary conditions and pedestal structure.

The approach is particularly well suited for modeling ELM-free regimes, such as the broad-pedestal lithiated scenarios of interest in NSTX. In future work, the same methodology could be extended to the prediction of density profiles, provided that reasonable estimates of the particle source are available.

\section*{Acknowledgements}
The authors would like to thank Emiliano Fable and Giovanni Tardini for their valuable assistance and guidance with the \textsc{astra} transport solver.  This research used resources of the National Energy Research Scientific Computing Center, a DOE Office of Science User Facility. This work was supported by U.S. Department of Energy Contract numbers: DE-SC0022115, DE-FG02-04ER54742, DE-SC0024425.


\vspace{1em}

\begin{thebibliography}{99}

\bibitem{Doyle2007}
E.~J.~Doyle \textit{et al.},
\emph{Nucl. Fusion} \textbf{47}, S18 (2007).

\bibitem{Howard2016}
N.~T.~Howard, A.~E.~White, M.~Greenwald, M.~Reinke, and J.~Candy,
\emph{Phys. Plasmas} \textbf{23}, 056109 (2016).

\bibitem{Parisi2024}
J.~F.~Parisi \textit{et al.},
\emph{Nucl. Fusion} \textbf{64}, 054002 (2024).

\bibitem{Horton1999}
W.~Horton,
\emph{Rev. Mod. Phys.} \textbf{71}, 735 (1999).

\bibitem{Jenko2000}
F.~Jenko, W.~Dorland, M.~Kotschenreuther, and B.~N.~Rogers,
\emph{Phys. Plasmas} \textbf{7}, 1904 (2000).

\bibitem{Dannert2005}
T.~Dannert and F.~Jenko,
\emph{Phys. Plasmas} \textbf{12}, 072309 (2005).

\bibitem{Jenko2005}
F.~Jenko \textit{et al.},
\emph{Plasma Phys. Control. Fusion} \textbf{47}, B195 (2005).

\bibitem{Hatch2017}
D.~R.~Hatch, M.~T.~Kotschenreuther, S.~M.~Mahajan, P.-Y.~Li, and R.~E.~Waltz,
\emph{Nucl. Fusion} \textbf{57}, 036020 (2017).

\bibitem{Kotschenreuther2019}
M.~T.~Kotschenreuther, D.~R.~Hatch, P.-Y.~Li, R.~Groebner, and S.~M.~Mahajan,
\emph{Nucl. Fusion} \textbf{59}, 096001 (2019).

\bibitem{Li2020}
P.-Y.~Li, D.~R.~Hatch, and M.~T.~Kotschenreuther,
\emph{Nucl. Fusion} \textbf{60}, 066007 (2020).

\bibitem{Groebner2022}
R.~J.~Groebner, D.~R.~Hatch, M.~E.~Fenstermacher, and P.~B.~Snyder,
\emph{Nucl. Fusion} \textbf{62}, 126022 (2022).

\bibitem{Parisi2023}
J.~Parisi, D.~R.~Hatch, N.~T.~Howard, and R.~E.~Waltz,
\emph{Nucl. Fusion} \textbf{63}, 046014 (2023).

\bibitem{Guttenfelder2012}
W.~Guttenfelder \textit{et al.},
\emph{Phys. Plasmas} \textbf{19}, 056119 (2012).

\bibitem{Diallo2013}
A.~Diallo \textit{et al.},
\emph{Nucl. Fusion} \textbf{53}, 093026 (2013).

\bibitem{Kaye2013}
S.~M.~Kaye \textit{et al.}, 
\emph{Nucl. Fusion} \textbf{53}, 063005 (2013).

\bibitem{Gorler2011}
T.~Görler, X.~Lapillonne, S.~Brunner, T.~Dannert, F.~Jenko, F.~Merz, and D.~Told,
\emph{J. Comput. Phys.} \textbf{230}, 7053 (2011).

\bibitem{Pereverzev2002}
G.~V.~Pereverzev and P.~N.~Yushmanov,
\emph{ASTRA Automated System for TRansport Analysis in a Tokamak}, Max-Planck-Institut für Plasmaphysik Report IPP 5/98 (2002).

\bibitem{Fable2013}
E.~Fable \textit{et al.},
\emph{Plasma Phys. Control. Fusion} \textbf{55}, 124028 (2013).

\bibitem{Tardini2026}
G.~Tardini \textit{et al.}, submitted to \emph{Plasma Phys. Control. Fusion} (2026).

\bibitem{Hatch2024}
D.~R.~Hatch, M.~T.~Kotschenreuther, P.-Y.~Li, B.~Chapman-Oplopoiou, J.~Parisi, S.~M.~Mahajan, and R.~Groebner,
\emph{Nuclear Fusion}, vol.~64, no.~6, 066007 (2024).
doi:10.1088/1741-4326/ad3972.

\bibitem{Candy2016}
J.~Candy, E.~A.~Belli, and R.~V.~Bravenec,
\emph{J. Comput. Phys.} \textbf{324}, 73 (2016).

\bibitem{Ivanov2005}
A.~A.~Ivanov \textit{et al.},
\emph{Non-linear 2D MHD equilibrium code SPIDER}, 32nd EPS Conference on Plasma Phys. (2005).

\bibitem{Lutjens1996}
H.~Lütjens, A.~Bondeson, and O.~Sauter,
\emph{Comput. Phys. Commun.} \textbf{97}, 219 (1996).

\bibitem{Kennedy2024}
D.~Kennedy \textit{et al.},
\emph{Nucl. Fusion} \textbf{64}, 016008 (2024).

\bibitem{Giacomin2025}
M.~Giacomin \textit{et al.},
\emph{J. Plasma Phys.} \textbf{91}, 905910103 (2025).

\bibitem{Hatch2018}
D.~R.~Hatch, R.~D.~Hazeltine, M.~K.~Kotschenreuther, and S.~M.~Mahajan,
\emph{Plasma Phys. Control. Fusion} \textbf{60}, 084003 (2018).
doi:10.1088/1361-6587/aac7a7.

\bibitem{Candy2003}
J.~Candy and R.~E.~Waltz,
\emph{J. Comput. Phys.} \textbf{186}, 545 (2003).

\end{thebibliography}
\end{document}